\documentclass[manuscript]{aastex}

\usepackage{color}
\usepackage{amsmath}
\usepackage{lscape}
\usepackage{graphicx}

\newcommand{\mr}{\mathrm} %Roman type

\newcommand{\sgmg}{\Sigma_\mr{gas}} %surface gas density
\newcommand{\rmol}{R_\mr{mol}} %rmol
\newcommand{\sgms}{\Sigma_\mr{star}} %surface stellar density
\newcommand{\sgmsf}{\Sigma_\mr{SFR}} %surface star formation rate density
\newcommand{\sgmin}{\Sigma_\mr{in}} %surface inflow rate density
\newcommand{\sgmout}{\Sigma_\mr{out}} %surface outflow rate density
\newcommand{\ret}{\mathcal{R}} %Return fraction
\newcommand{\tr}{(t,\:R)} %(t, R)
\newcommand{\rf}{R_\mr{f}} %R_f
\newcommand{\tf}{t_\mr{f}} %t_f
\newcommand{\srm}{\sigma_\mr{RM}} %sgm_RM
\newcommand{\trm}{t_\mr{RM}} %t_RM
\newcommand{\srmz}{\sigma_\mr{RM,0}} %sgm_RM

\newcommand{\mtotin}{M_\mr{tot,in}} %M_tot,in
\newcommand{\hrin}{h_\mr{R,in}} %h_r,in
\newcommand{\tauin}{\tau_\mr{in}} %tau_in

\newcommand{\tauinz}{\tau_\mr{in,0}} %tau_in,0
\newcommand{\tauine}{\tau_\mr{in,8}} %tau_in,8

\newcommand{\fia}{f_\mr{Ia}} %f_Ia

\shortauthors{TOYOUCHI \& CHIBA.}
\shorttitle{Bimodal distribution of Galactic disk stars on the [$\alpha$/Fe]-[Fe/H] plane as a possible evidence of discontinuous radial migration history}

\begin{document}

\title{Bimodal Distribution of Galactic Disk Stars on the [$\alpha$/Fe]-[Fe/H] Plane as a Possible Evidence of Discontinuous Radial Migration History}

\author{Daisuke~Toyouchi\altaffilmark{1} and
	Masashi~Chiba\altaffilmark{1}}

\altaffiltext{1}{Astronomical Institute, Tohoku University,
Aoba-ku, Sendai 980-8578, Japan}

%%%%%% Abstract %%%%%%%%%%%%%%%%%%%%%%%%%%%%%%%%%%%%%%%%%%
\begin{abstract}
We investigate the role of radial migration history of stars in chemical evolution of a disk galaxy, in particular in understanding the origin of their bimodal distribution on the [$\alpha$/Fe]-[Fe/H] plane. For this purpose, we examine the three different models with no, continuous, and discontinuous radial migration, respectively. We find that for the model with radial migration, the [$\alpha$/Fe] ratios of stars in outer disk regions decrease more rapidly with time than the model without radial migration, because the associated net transfer of intermediate and old disk stars from inner to outer disk regions increases the rate of Type Ia relative to that of Type II supernovae in the latter regions. Moreover, in the model assuming rapid and discontinuous radial migration, its effect on the stellar abundances at larger radii is significant enough to provide the large difference in the evolution of stars on the [$\alpha$/Fe]-[Fe/H] plane between inner and outer disk regions. As a result we obtain the bimodal distribution of disk stars on the [$\alpha$/Fe]-[Fe/H] plane as observed in the Galactic stellar disk, thereby implying that the event of discontinuous radial migration may play a key role in reproducing the observed bimodality of stars on this abundance-ratio diagram. We discuss possible mechanisms causing such discontinuous radial migration in the early evolution of the Galactic disk, including the event of minor merging of a relatively massive satellite onto the stellar disk.  

\end{abstract}
%%%%%%%%%%%%%%%%%%%%%%%%%%%%%%%%%%%%%%%%%%%%%%%%%%%%%%%%%%


%%% Sec.1 %%%%%%%%%%%%%%%%%%%%%%%%%%%%%%%%%%%%%%%%%%%%%%%%
\section{INTRODUCTION}
The Milky Way galaxy is the best site for understanding the formation and evolution processes of disk galaxies, because we can observe individual stars composing the stellar disk in great detail. Many observations of the Galactic disk stars have shown the various detailed properties of the present stellar disk, such as the spatial structure (e.g., Yoshii 1982; Gilmore \& Reid 1983; Juri\' c et al. 2008; Bovy et al. 2012, 2016), metallicity distribution (e.g., Wyse \& Gilmore 1995; Lee et al. 2011; Hayden et al. 2015), and radial metallicity gradient (e.g., Nordstr\"om et al. 2004; Allende Prieto et al. 2006; Cheng et al. 2012; Toyouchi \& Chiba 2014). However there are still many mysteries in the formation history of the Galactic disk (see Feltzing \& Chiba 2013; Rix \& Bovy 2013 for reviews). 

One of the mysteries is the bimodal distribution of the Galactic disk stars on the [$\alpha$/Fe]-[Fe/H] plane: the high and low-$\alpha$ peaks of stellar density on the chemical abundance plane appear around the [$\alpha$/Fe] ratio of  $\sim$ 0.2-0.3 and $\sim$ 0, respectively (e.g., Bensby et al. 2003; Lee et al. 2011; Adibekyan et al. 2012; Anders et al. 2014; Mikolaitis et al. 2014). Previous observations have shown that the high-$\alpha$ sequence stars belong to a dynamically hotter and geometrically thicker disk component than the low-$\alpha$ sequence stars (e.g., Bensby et al. 2014; Recio-Blanco et al. 2014). Therefore some studies suggested that the definition of the thick and thin disk stars should be based on the chemical abundances of the disk stars (e.g., Navarro et al. 2011; Lee et al. 2011), although the existence of actually distinct thick and thin disk components in our Galaxy is still a matter of debate (e.g., Bovy et al. 2012).

How the bimodal distribution of the disk stars on the [$\alpha$/Fe]-[Fe/H] plane are formed in the course of the Galactic disk formation is not clearly understood. A possible scenario to reproduce such a bimodal distribution is a brief cessation of star formation at the early disk formation phase (e.g., Chiappini et al. 1997, 2001). In this scenario, after the early intense disk formation phase, the star formation activity in the disk temporarily ceases. Such a cessation of star formation leads a rapid decrease of [$\alpha$/Fe] ratio of interstellar medium because chemical enrichment is mostly dominated by Type Ia supernovae (SNe), not Type II SNe. As a result, the disk stars born before and after this cessation of star formation have distinctly different [$\alpha$/Fe] ratios, respectively, thereby reproducing the bimodality on the [$\alpha$/Fe]-[Fe/H] plane. Recently, Haywood et al. (2016) showed based on the simple closed box chemical evolution model that the significant drop of star formation rate from 10 to 8 Gyr ago is an essential ingredient to reproduce the bimodal distribution of the Galactic disk stars on the [$\alpha$/Fe]-[Fe/H] plane. However, realizing this scenario requires a specific change of star formation efficiency, for which it may be difficult to find an appropriate physical interpretation. Thus, it is worth exploring yet another idea to explain the existence of such a bimodal distribution.

One of the important processes in disk evolution is radial migration of disk stars, generally driven by bar/spiral structures in the galactic disk (e.g., Sellwood \& Binney 2002; Ro{\v s}kar et al. 2008; Loebman et al. 2011). Hayden et al. (2015) showed that angular momentum redistribution processes in the stellar disk are necessary to interpret the radial dependence of the metallicity distribution of the Galactic disk stars revealed by the observation of SDSS-III/APOGEE. Sch\"onrich \& Binney (2009a, b) constructed the semi-analytic chemo-dynamical evolution model including such an effect of radial migration, and successfully reproduced the metallicity distribution and age-metallicity relation of the disk stars in the solar neighborhood. Moreover, they suggested that the bimodal distribution on the [$\alpha$/Fe]-[Fe/H] plane is naturally understood by the effect of radial migration of stars driven by bar/spiral structures in the galactic disk. According to their model calculation, the high-$\alpha$ sequence consists of the old disk stars, which have radially migrated from the inner disk regions, whereas most of the stars belonging to the low-$\alpha$ sequence are born around the solar annulus. However, the recent chemo-dynamical evolution model of Minchev et al. (2013), in which the chemical and dynamical evolution were independently calculated by the semi-analytical model and the cosmological numerical simulation, respectively, showed that radial migration alone is insufficient to construct the observed bimodal distribution on the [$\alpha$/Fe]-[Fe/H] plane.

In addition to the internal process associated with bar/spiral structures, the external perturbation such as minor merging of a satellite galaxy and subsequent dynamical heating of the galactic disk can also drive radial migration by depositing the orbital energy and angular momentum of the merging satellite into the galactic disk stars (e.g., Quinn et al. 1993; Vel\'{a}zquez \& White 1999; Villalobos \& Helmi 2008). The radial redistribution of stars triggered by a minor merger is a more dramatical and discontinuous event than that induced by bar/spiral structures. Minor merger events, possibly leading such rapid radial expansions of the stellar disk, are expected to commonly occur at an early disk formation phase (e.g., Ruiz-Lara et al. 2016). Indeed, Minchev et al. (2013) pointed out the importance of minor merger heating events to thicken the stellar disk up to the same level as the observed Galactic oldest disk component. However, the impact of such a discontinuous radial migration process on the chemical evolution of the galactic disk remains to be understood.

In this study, we reconsider the effect of radial migration of disk stars as an key ingredient to reproduce the bimodal distribution on the [$\alpha$/Fe]-[Fe/H] plane based on the calculation of a semi-analytical chemo-dynamical evolution model. We particularly focus on the influence of the different radial migration histories on the chemical evolution of the disk galaxy, and for this purpose we consider not only continuous radial migration as driven by bar/spiral structures, but also discontinuous one as driven by minor merger of a satellite. We will show below that radial migration of disk stars is an important process to affect the chemical evolution over the galactic disk and that the incident of discontinuous radial migration process may be essential to reproduce the observed bimodal distribution of the disk stars on the [$\alpha$/Fe]-[Fe/H] plane.

This paper is organized as follows. In Section 2, we introduce our chemo-dynamical evolution model. In Section 3, we show the stellar distribution on the [$\alpha$/Fe]-[Fe/H] plane obtained from our model calculations and its dependence on the assumed radial migration histories. In Section 4, we discuss the dependence of the results on model parameters and suggest the specific radial migration history for the stellar disk of the Milky Way. Finally, our conclusions are drawn in Section 5.

%%% Sec.2 %%%%%%%%%%%%%%%%%%%%%%%%%%%%%%%%%%%%%%%%%%%%%%%%
\section{MODEL}
To investigate the influence of radial migration on a chemical enrichment history of a disk galaxy, we adopt a standard chemo-dynamical model, which has been studied in many previous works (e.g., Sch{\"o}nrich \& Binney 2009a; Kubryk et al. 2015). In this model, a galactic disk consists of many co-center rings, and by calculating baryonic mass evolution for each ring at each time step, we obtain the surface density of gas, $\sgmg$, that of stars, $\sgms$, and a mass fraction of heavy element $i$, $Z_i$, respectively, at any time, $t$, and at any radius, $R$. These calculations are carried out in a radial range from $R$ = 0 to $R_\mr{out}$ (= 16 kpc) with a grid of $\Delta R$ = 1 kpc over $t$ = 0 to $t_\mr{p}$ (= 12 Gyr) with a grid of $\Delta t$ = 50 Myr. In this paper, we are particularly interested in the evolution of [$\alpha$/Fe], which is defined here as the average for the typical $\alpha$ elements of O, Mg, Si, Ca, and Ti.

In Section 2.1-2.4, we introduce each important process in our chemical evolution model. The main equations to calculate baryonic mass evolution on a galactic disk are described in Section 2.5. Finally, in Section 2.6 we introduce how to choose the values of the important free parameters in our model.

%%% Sec.2.1 %%%%%%%%%%%%%%%%%%%%%%%%%%%%%%%%%%%%%%%%%%%%%%%%
\subsection{Gas Inflow Rate}
A galactic disk is thought to form from an inflowing gas from outside in the course of galaxy formation.  In our model, we assume that the surface density of gas inflow rate, $\sgmin$, as functions of $t$ and $R$ is given as,
\begin{eqnarray}
\sgmin \tr \ = \  \frac{\mtotin}{2 \pi \hrin^2} \frac{ \mr{exp} \left ( -R/{\hrin} - t/{\tauin} \right )}{\tauin \left[1- \mr{exp}(-t_\mr{p}/\tauin) \right]} \ ,
\label{eq:inflow}
\end{eqnarray}
where $\mtotin$ and $\hrin$ are the total mass and scale length of inflowing gas on the disk plane, respectively. $\tauin$ is a time scale of gas inflow rate at $R$, described with $\tauinz$, $\tauine$ and $\alpha$ in our model as,
\begin{eqnarray}
\tauin (R) = \tauinz + (\tauine-\tauinz) \left ( \frac{R}{8 \ \mr{kpc}} \right )^{\alpha} \ .
\label{eq:tauin}
\end{eqnarray}
This model for gas inflow rate has five parameters ($\mtotin$, $\hrin$, $\tauinz$, $\tauine$, $\alpha$).

%%% Sec.2.2 %%%%%%%%%%%%%%%%%%%%%%%%%%%%%%%%%%%%%%%%%%%%%%%%
\subsection{Star Formation Rate}
The surface density of star formation rate, $\sgmsf$, in our model follows the observationally motivated star formation law by Bigiel et al. (2008), in which $\sgmsf$ is proportional to the surface density of H$_2$ gas. The star formation rate can be written using the mass ratio of H$_2$ to HI gas, $\rmol$, as follows,
\begin{eqnarray}
\sgmsf = 1.6 \  \frac{\rmol}{\rmol+1} \  \left ( \frac{\sgmg}{M_\odot \mr{pc}^{-2}} \right ) \ \ \ [M_\odot \ \mr{pc}^{-2} \ \mr{Gyr}^{-1}] \ .
\label{eq:sfr}
\end{eqnarray}
To calculate $\sgmsf$ from this equation, we additionally make use of the semi-empirical law of $\rmol$ provided by Blitz \& Rosolowsky (2006), which depends on $\sgmg$ and $\sgms$ as follows,
\begin{eqnarray}
\rmol = 0.23 \ \left [ \left ( \frac{\sgmg}{10 \ M_\odot \mr{pc}^{-2}} \right ) \left ( \frac{\sgms}{35 \ M_\odot \mr{pc}^{-2}} \right )^{0.5} \right ]^{0.92} \ .
\label{eq:rmol}
\end{eqnarray}
These empirical laws of equations (\ref{eq:sfr}) and (\ref{eq:rmol}) have been known to reproduce the present spatial distribution of HI, H$_2$ and star formation in the Milky Way (Blitz \& Rosolowsky 2006; Kubryk et al. 2015).

%%% Sec.2.3 %%%%%%%%%%%%%%%%%%%%%%%%%%%%%%%%%%%%%%%%%%%%%%%%
\subsection{Gas Outflow Rate}
Feedback processes associated with star formation activity, such as radiation pressure from massive stars and supernova explosions, may drive a strong galactic gas outflow, which gives a significant impact on the chemical evolution of a galactic disk. In this model, we assume that the surface density of gas outflow rate, $\sgmout$, is proportional to $\sgmsf$ with an assumed outflow-mass loading factor, $\Lambda$, which is described as a following function,
\begin{eqnarray}
\Lambda(R) = \Lambda_0 + (\Lambda_8-\Lambda_0) \left ( \frac{R}{8 \ \mr{kpc}} \right )^{\beta} \ ,
\label{eq:outflow}
\end{eqnarray}
where $\Lambda_0$, $\Lambda_8$ and $\beta$ are parameters characterizing the radial dependence of $\Lambda$.

It is worth noting that this expression of gas outflow tightly linking with star formation approximately includes the effect of gas radial flow along a galactic disk, because the radial gas flow is expected to be driven by gravitational torques related to the formation of spiral/bar or giant molecular clouds, which are also important drivers of star formation, so that radial gas flow may occur with a time scales similar to that of star formation (Yoshii \& Sommer-Larsen 1989). Therefore, in our model $\Lambda$ can be negative, when the amount of gas expelled from the disk region as outflow is less than that of gas supplied into the region by radial gas flow, although $\Lambda$ calculated in Section 2.6 are found to be positive at all radii.

We also note that while for simplicity we assume that $\Lambda$ does not evolve with time, this assumption may not be appropriate. According to previous works on chemical evolution models applied to several observational results of extra-galactic star-forming galaxies, the outflow-mass loading factor increases with increasing redshift (e.g. Yabe et al. 2015; Toyouchi \& Chiba 2015). Therefore, our time-independent mass loading factor may be regarded as an time-averaged one, although it actually changes with time.

%%% Sec.2.4 %%%%%%%%%%%%%%%%%%%%%%%%%%%%%%%%%%%%%%%%%%%%%%%%
\subsection{Radial migration}
To take into account the effect of radial migration of disk stars in our model, we make use of the method adopted in Sellwood \& Binney (2002), which can reproduce the basic properties of radial migration obtained in N-body simulations well. This method provides the probability, in which a star born at radius $\rf$ and time $\tf$ is found later in radius $R$ and time $t$, $P(t, \ R, \ \tf, \ \rf)$, and this is expressed in terms of the following Gaussian function 
\begin{eqnarray}
P(t, \ R, \ \tf, \ \rf) =  \frac{1}{\sqrt{2 \pi \srm^2}} \ \mr{exp} \left [ \ -\frac{(R-\rf)^2}{2 \srm^2} \ \right ] \ ,
\label{eq:p_rm}
\end{eqnarray}
where $\srm$ corresponds to the diffusion length of stars by radial migration, which is generally a function of $t$, $\tf$, and $\rf$. Thus, in this method the radial migration history can be characterized by the time dependence of $\srm$. 

\begin{figure}
\begin{center}
\includegraphics[width=12cm,height=18cm]{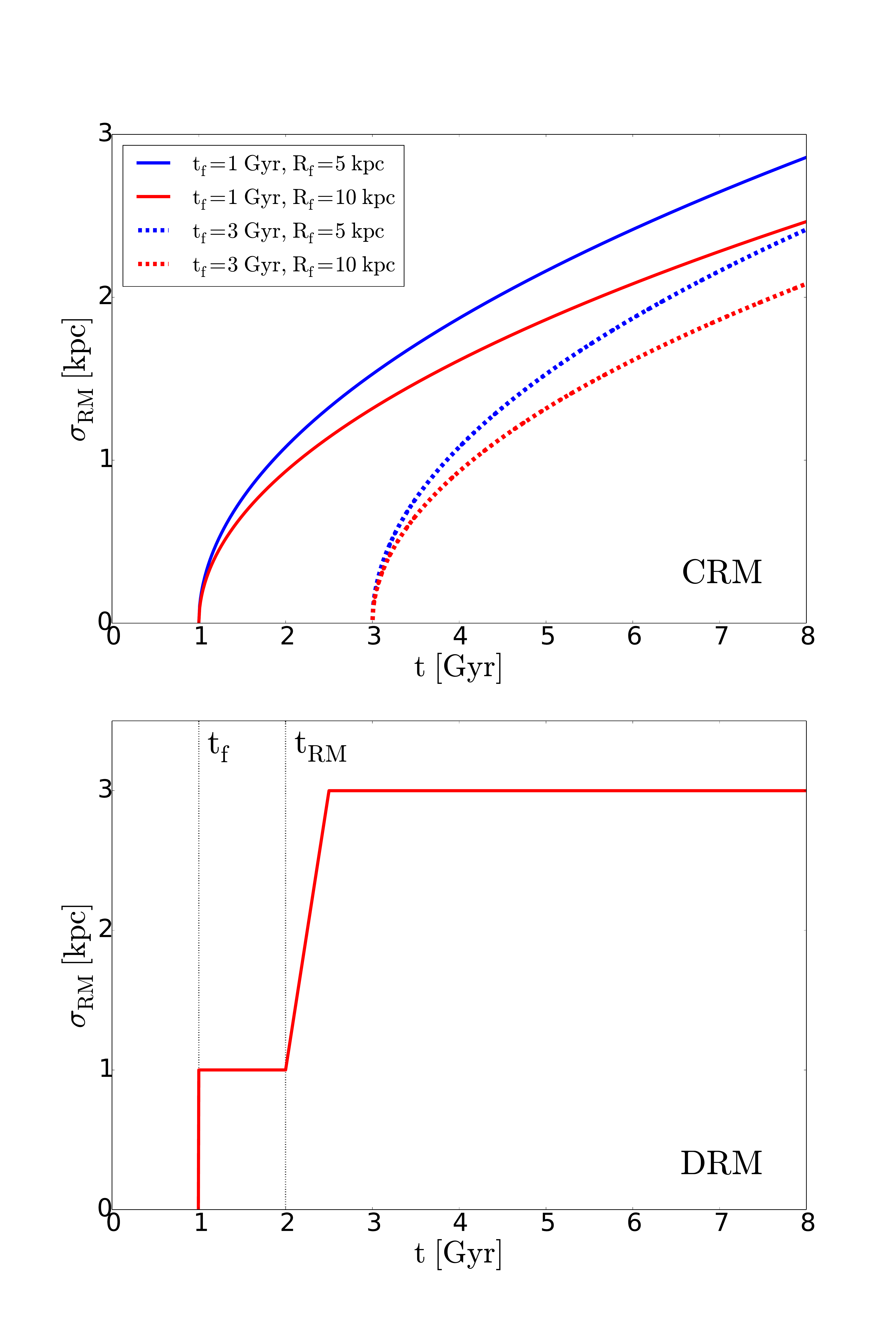}
\end{center}
\caption{The upper panel shows the evolution of $\srm$ in the CRM model for the cases of $\tf$ = 1 Gyr (solid) and 3 Gyr (dashed), and $\rf$ = 5 kpc (blue) and 10 kpc (red). The lower panel represents the evolution of $\srm$ for stars born at $t < \trm$ in the DRM model. In this panel we set $\tf$ = 1 Gyr, $\trm$ = 2 Gyr, and $\srmz$ = 3 kpc. The radial migrations of stars born at $t > \trm$ in the DRM model follow the same scheme as the CRM model as shown in the upper panel.}
\label{fig:srm} 
\end{figure}

In our study, to investigate how the difference in radial migration histories affects the chemical evolution of a galactic disk, we consider three different models for $\srm$. The first model does not include the effect of radial migration, hereafter no radial migration (NRM) model, where $\srm$ is always zero. Second is the continuous radial migration (CRM) model, where $\srm$ monotonically increases with increasing $t-\tf$, which corresponds to the stellar age, as follows,
\begin{eqnarray}
\srm(t, \ \tf, \ \rf) = (-0.0667 \rf + 2.75) \left ( \frac{t-\tf}{5 \ \mr{Gyr}} \right )^{0.5} \ \ [ \mr{kpc} ] \ .
\label{eq:crm}
\end{eqnarray}
The evolution of $\srm$ described by this equation is demonstrated in the upper panel of Figure \ref{fig:srm}, where we show the cases of ($\tf$/Gyr, $\rf$/kpc)  = (1, 5),  (1, 10), (3, 5) and (3, 10). Equation (\ref{eq:crm}) is based on the result of Kubryk et al. (2013), who analyzed the radial migration history in a bar-dominated disk galaxy in the high-resolution N-body+smoothed particle hydrodynamics simulation. The properties of the radial migration in equation (\ref{eq:crm}) has been known to be roughly similar to that in the chemo-dynamical model of Sch\"onrich \& Binney (2009a), who adopt a different scheme to represent radial migration (Kubryk et al. 2015). Therefore, we study the CRM model as an appropriate test case to examine the influence of continuous radial migration driven by internal mechanisms, such as interaction between bar/spiral and stars, on the chemical evolution of a galactic disk.

Our third model is the discontinuous radial migration (DRM) model, where $\srm$ evolves discontinuously at $t$ = $\trm$ as follows, 
\begin{eqnarray}
\srm(t, \ \tf, \ \rf) = \begin{cases}
\scalebox{1.0}{$\displaystyle \ \ \ \ 1 \ \ [\mr{kpc}]$} & (t \leq \trm) \\[11pt]
\scalebox{1.0}{$\displaystyle (\srmz-1)  \times \mr{min} \left [ 1, \ \left ( \frac{t-\tf}{0.5 \ \mr{Gyr}} \right ) \right ]+1  \ \ [\mr{kpc}]$} & (t > \trm \ \& \ \tf \leq \trm) \\[11pt]
\scalebox{1.0}{$\displaystyle (-0.0667 \rf + 2.75) \left ( \frac{t-\tf}{5 \ \mr{Gyr}} \right )^{0.5}  \ \ [\mr{kpc}]$} & (t > \trm \ \& \ \tf > \trm) \ .
\end{cases}
\label{eq:drm}
\end{eqnarray}
In this model, we assume that while at $t < \trm$, the disk stars have experienced only moderate radial migration with radial diffusion length of 1 kpc, which is roughly consistent with the scale length of the Milky Way-like galaxies at the early disk formation phase. At $t = \trm$ the rapid radial migration is assumed to occur with the radial diffusion length, $\srmz$, in which the time scale of the diffusion is 0.5 Gyr, corresponding to the several dynamical times of the Milky Way disk with the circular velocity of $\sim$ 200 km s$^{-1}$. This evolution of $\srm$ for $\tf < \trm$ in the DRM model is demonstrated in the lower panel of Figure \ref{fig:srm}, where we show the case of $\tf$ = 1 Gyr, $\trm$ = 2 Gyr and $\srmz$ = 3 kpc. Subsequently the disk stars, which were born at $t > \trm$, have followed the same radial migration history as the CRM model. In this DRM model, $\trm$ and $\srmz$ are key parameters characterizing the properties of the discontinuous radial migration event. As our fiducial DRM model, we choose $\srmz$ = 3 kpc, which is roughly consistent with the scale length of the present Milky Way disk, and $\trm$ = 2 Gyr, corresponding to the look back time of 10 Gyr, which is in roughly agreement with the specific time when the age-metallicity and age-[$\alpha$/Fe] relation of the Milky Way disk stars in the solar neighborhood dramatically evolve (Haywood et al. 2013). How the choice of the values of $\srmz$ and $\trm$ affects our results will be discussed in Section 4.1. 

We note here that such rapid and discontinuous radial migration in the DRM model is supposed to be triggered by an external perturbation, such as minor merger heating, rather than the internal processes as considered in the CRM model. Thus, by comparing these three radial migration models, we explore the impact on the chemical evolution of the galactic disk by both internal and external dynamical heating events.

%%% Sec.2.5 %%%%%%%%%%%%%%%%%%%%%%%%%%%%%%%%%%%%%%%%%%%%%%%%
\subsection{Basic Equations}
To obtain the time evolution of $\sgmg$, $\sgms$ and $Z_i$, we solve the following equations, including the effects of star formation, gas inflow, gas outflow and radial migration of stars,
 \begin{eqnarray}
\frac{\partial  \sgmg}{\partial t} = - (1 - \ret) \sgmsf + \sgmin - \sgmout \ ,
\label{eq:gas}
\end{eqnarray}
 \begin{eqnarray}
 \sgms \tr = (1 - \ret) \int^{R_\mr{out}}_{0} \int^{t}_{0} \frac{\rf}{R} \ \sgmsf (\tf, \ \rf)  P(t, \ R, \ \tf, \ \rf)  \mr{d}\tf  \mr{d}\rf \ ,
\label{eq:star}
\end{eqnarray}
 \begin{eqnarray}
\frac{\partial  (Z_i \sgmg)}{\partial t} = (Y_{\mr{II},i} + Y_{\mr{Ia},i}) \sgmsf - Z_i (1 - \ret) \sgmsf + Z_{\mr{in},i} \sgmin - Z_{\mr{out},i} \sgmout \;.
\label{eq:metal}
\end{eqnarray}
In equation (\ref{eq:gas}) the first term on the right hand side represents the net gas consumption by star formation, where $\ret$ is the mass fraction returned back to interstellar medium (ISM) via stellar mass loss, and the description of this term is based on an instantaneous recycling approximation. In this paper we set $\ret$ = 0.45 corresponding to the Chabrier initial mass function (Leitner \& Kravtsov 2011).

Equation (\ref{eq:star}) implies that the time evolution of $\sgms$ at any radius depends on the past star formation and radial migration history over the galactic disk. $R_\mr{out}$ in equation (\ref{eq:star}) is the outer limit of $R$ in our calculation, where we set $R_\mr{out}$ = 16 kpc, i.e., much larger than the disk size of the Milky Way to avoid its effect on the calculation.

In equation (\ref{eq:metal}) the first term on the right hand side describes the supply of heavy element $i$ newly synthesized in stars, where $Y_{\mr{II},i}$ and $Y_{\mr{Ia},i}$ are the nucleosynthetic yields from Type II and Ia SNe (hereafter SN II and SN Ia), respectively. We assume that $Y_{\mr{II},i}$ is constant and adopt the SN II yield from Fran\c{c}ois et al. (2004). On the other hand, $Y_{\mr{Ia},i}$ changes with time and radius, reflecting the past  star formation and radial migration history as follows, 
\begin{eqnarray}
Y_{\mr{Ia},i} = \frac{(1 - \ret) \ m_{\mr{Ia},i} \ \fia}{\sgmsf \tr} \int^{R_\mr{out}}_{0} \int^{t-\Delta t_\mr{Ia}}_{0}  \frac{\rf}{R} \ \frac{\sgmsf (\tf, \ \rf)  P(t, \ R, \ \tf, \ \rf)}{t-\tf}  \mr{d}\tf  \mr{d}\rf \ ,
\label{eq:yia}
\end{eqnarray}
where $\fia$ is a free parameter controlling the SN Ia rate in the galactic disk, and $\Delta t_\mr{Ia}$ is a minimum delayed time of SN Ia. Here we set $\Delta t_\mr{Ia}$ = 0.5 Gyr suggested by Homma et al. (2015), which reproduced, with this value of $\Delta t_\mr{Ia}$, star formation histories and chemical evolutions of the Galactic dwarf galaxies self-consistently. Another SN Ia parameter, $m_{\mr{Ia},i}$, in equation (\ref{eq:yia}) is the released mass of heavy element $i$ per a Type Ia supernova, for which we adopt the SN Ia yield of W7 model in Iwamoto et al. (1999). 

The second term on the right hand side of equation (\ref{eq:metal}) is the mass of heavy elements finally locked up in stars. The third and fourth terms on the right hand side of equation (\ref{eq:metal}) denote the mass injection and ejection of heavy elements associated with inflow and outflow, respectively, where $Z_{\mr{in},i}$ and $Z_{\mr{out},i}$ are mass fractions of heavy element $i$ in inflowing and outflowing gas, respectively. In this model, we set inflowing gas abundance of [Fe/H] = -1.5, which is the typical metallicity of the Galactic halo stars (e.g., Carollo et al. 2010), and [$\alpha$/Fe] = 0.4 for each $\alpha$ element, roughly corresponding to the average abundance ratio of $Y_\mr{II,\alpha}$ to $Y_\mr{II, Fe}$. Additionally we assume that the metallicity of outflowing gas corresponds to that of the ISM, implying $Z_{\mr{out},i} = Z_i$.

%%% Sec.2.6 %%%%%%%%%%%%%%%%%%%%%%%%%%%%%%%%%%%%%%%%%%%%%%%%
\subsection{Determination of Model Parameters}
This chemical evolution model contains nine free parameters, which are summarized in Table 1. In this study, we adopt the MCMC method (Metropolis et al. 1953; Hastings 1970) to obtain the best set of these nine free parameters, which reproduce the observed radial profiles of gas, star, [O/H] and [Fe/H] in the Milky Way disk. In this model, following the results of several previous works, we adopt the Galactic stellar disk with the scale length of 2.3 kpc and the stellar density at $R$ = 8 kpc of 35 $\mr{M_\odot} \ \mr{pc}^{-2}$ (Flynn et al. 2006). For the total gas density profile, consisting of HI and H$_2$, in the Galactic disk, we adopt the work of Kubryk et al. (2015) shown in their Figure (A.2), and the radial profile of [O/H] and [Fe/H] in the disk is taken from the observation of Cepheids in Luck \& Lambert (2011). 

\begin{figure}
\begin{center}
\includegraphics[width=18cm,height=15cm]{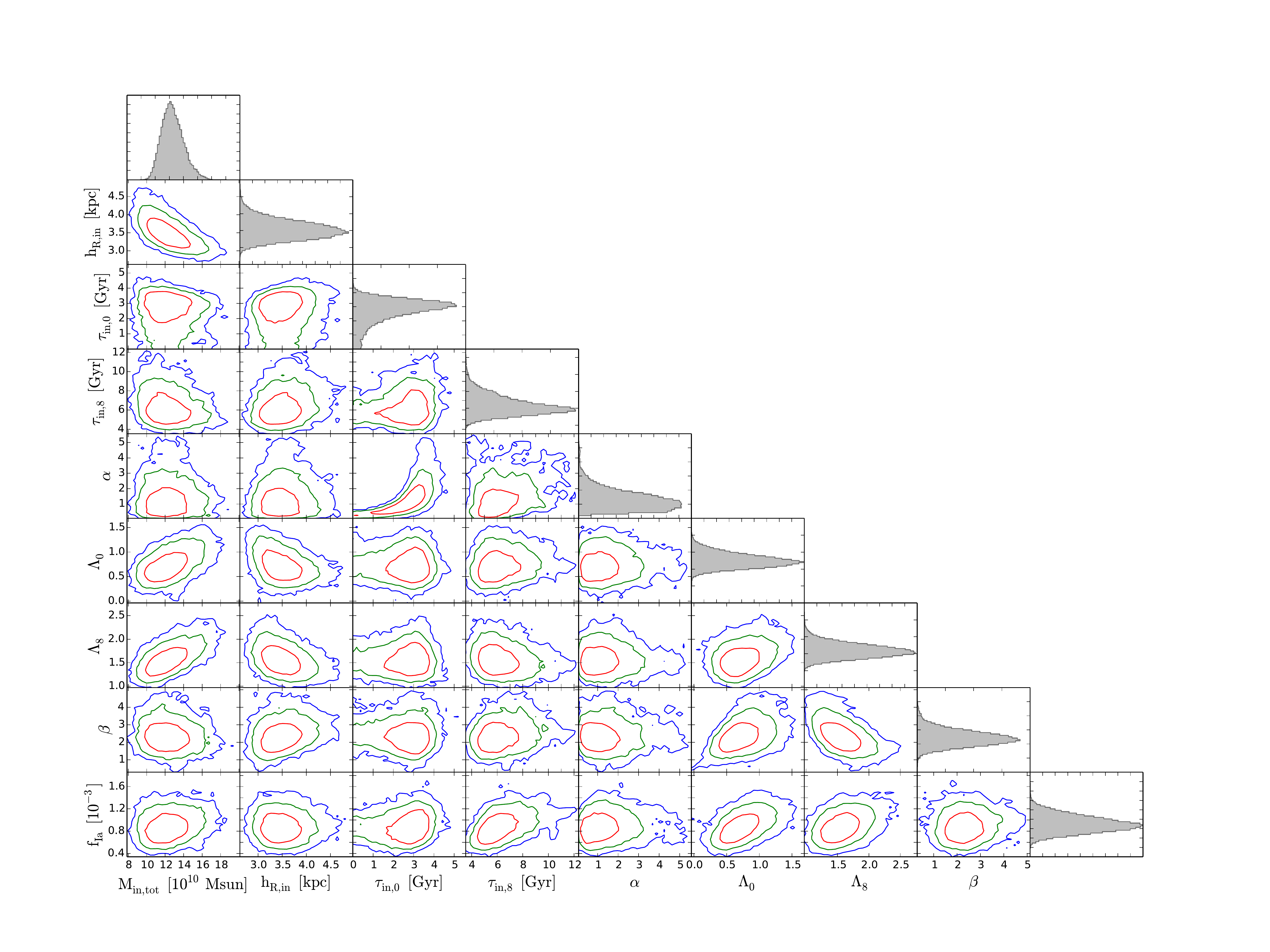}
\end{center}
\caption{The diagonal and off-diagonal panels show the 1D and 2D posterior probability distributions of the nine parameters for the NRM model, respectively. The red, green and blue contours in each off-diagonal panel represent the 1, 2 and 3 $\sigma$ regions of the posterior probability distributions, respectively.}
\label{fig:mc_nrm} 
\end{figure}

\begin{figure}
\begin{center}
\includegraphics[width=18cm,height=15cm]{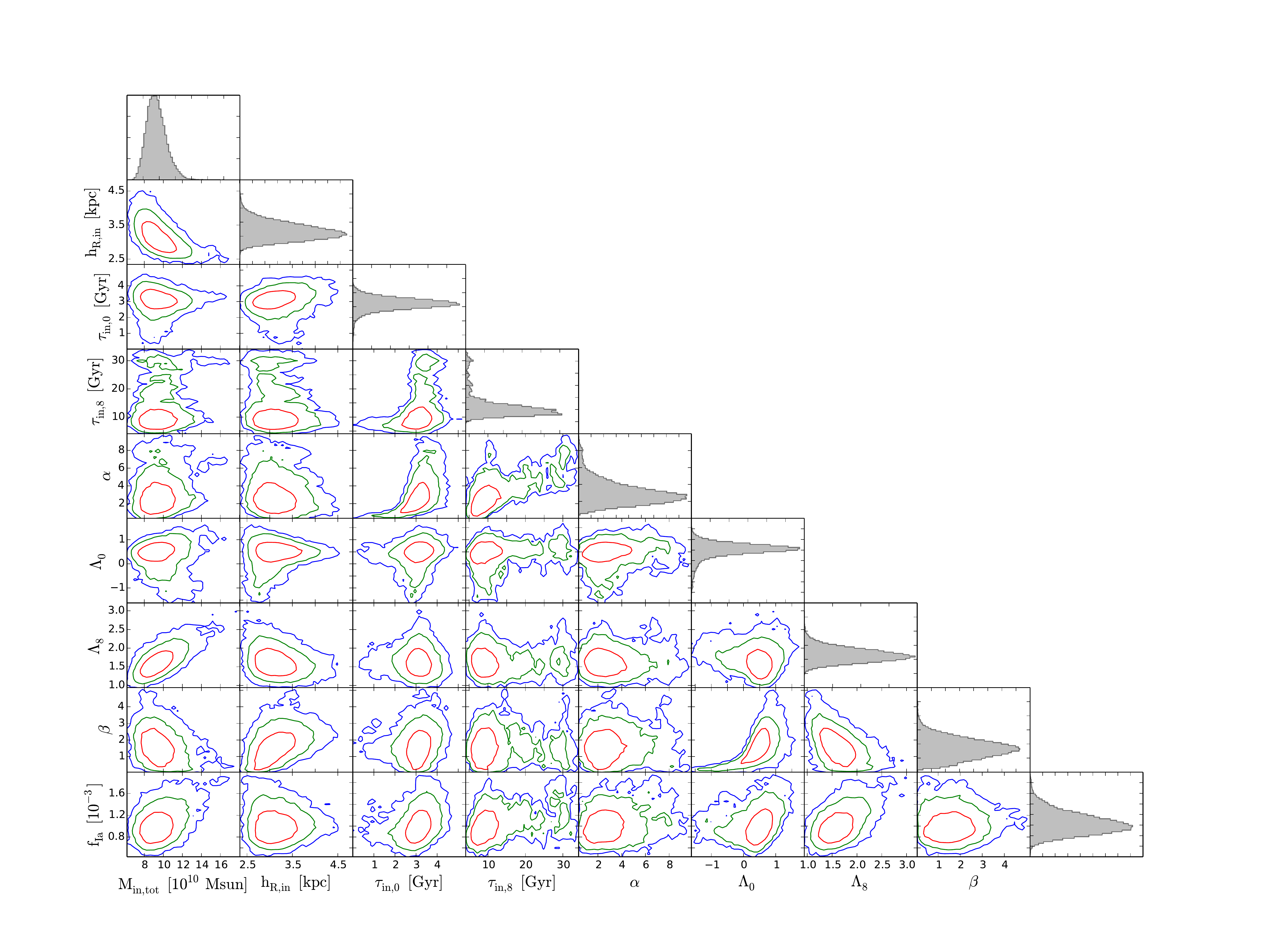}
\end{center}
\caption{The same as Figure \ref{fig:mc_nrm}, but for the CRM model.}
\label{fig:mc_crm} 
\end{figure}

\begin{figure}
\begin{center}
\includegraphics[width=18cm,height=15cm]{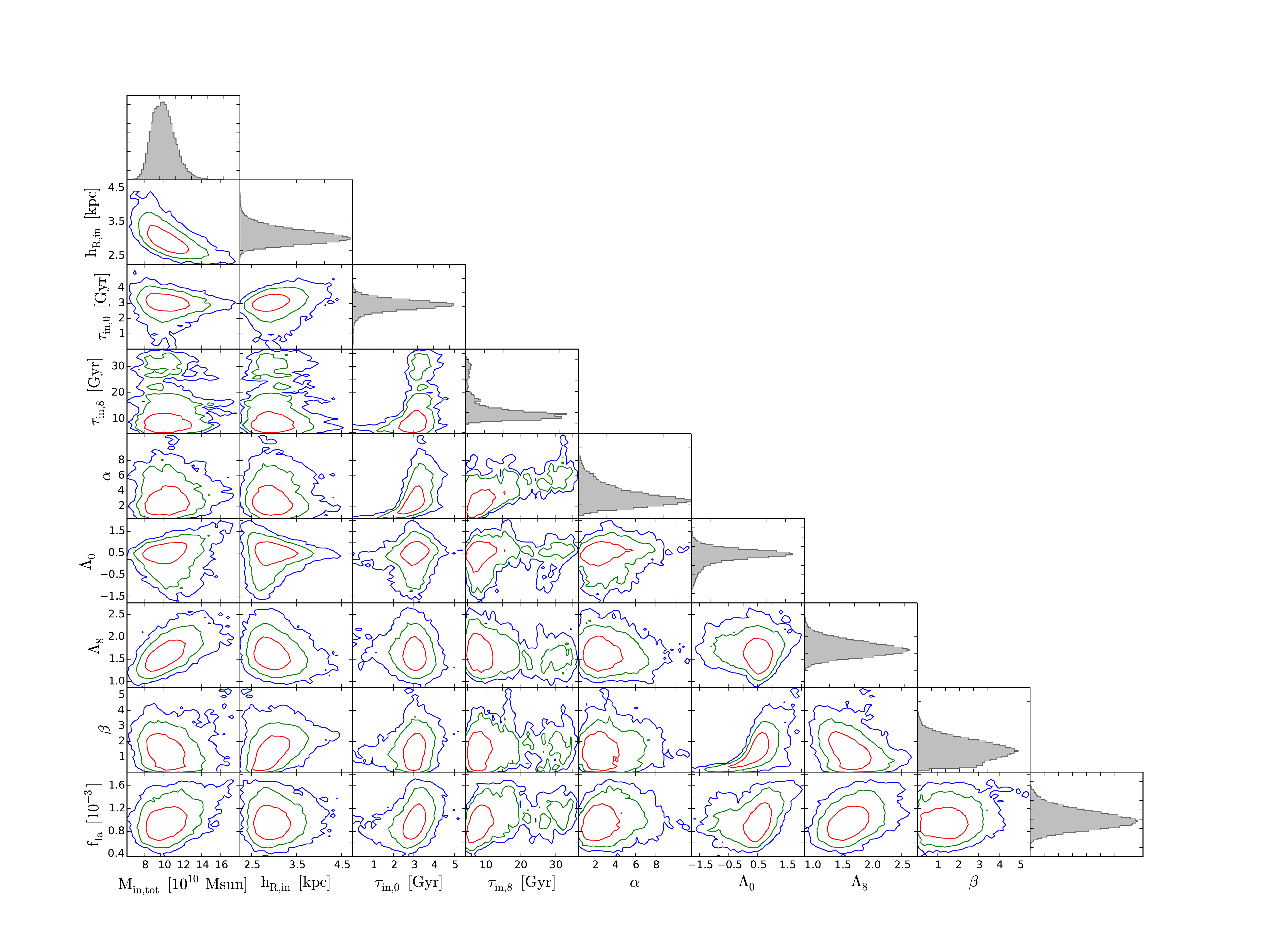}
\end{center}
\caption{The same as Figure \ref{fig:mc_nrm}, but for the DRM model.}
\label{fig:mc_drm} 
\end{figure}

In our MCMC procedure, for each radial migration model, we carry out 5 MCMC chains starting from different parameter sets. Each chain consists of 100,000 iterations, and by compiling the later 50,000 iterations in all chains, we get the posterior probability distribution for each parameter. Figure \ref{fig:mc_nrm}, \ref{fig:mc_crm} and \ref{fig:mc_drm} show the posterior probability distributions of the nine parameters for the NRM, CRM, and DRM models, respectively. We find from these figures that for all radial migration models, our MCMC chains for all parameters converge successfully. The results of this estimation of the nine parameters for each radial migration model are summarized in Table 2. 

\begin{figure}
\begin{center}
\includegraphics[width=17cm,height=12cm]{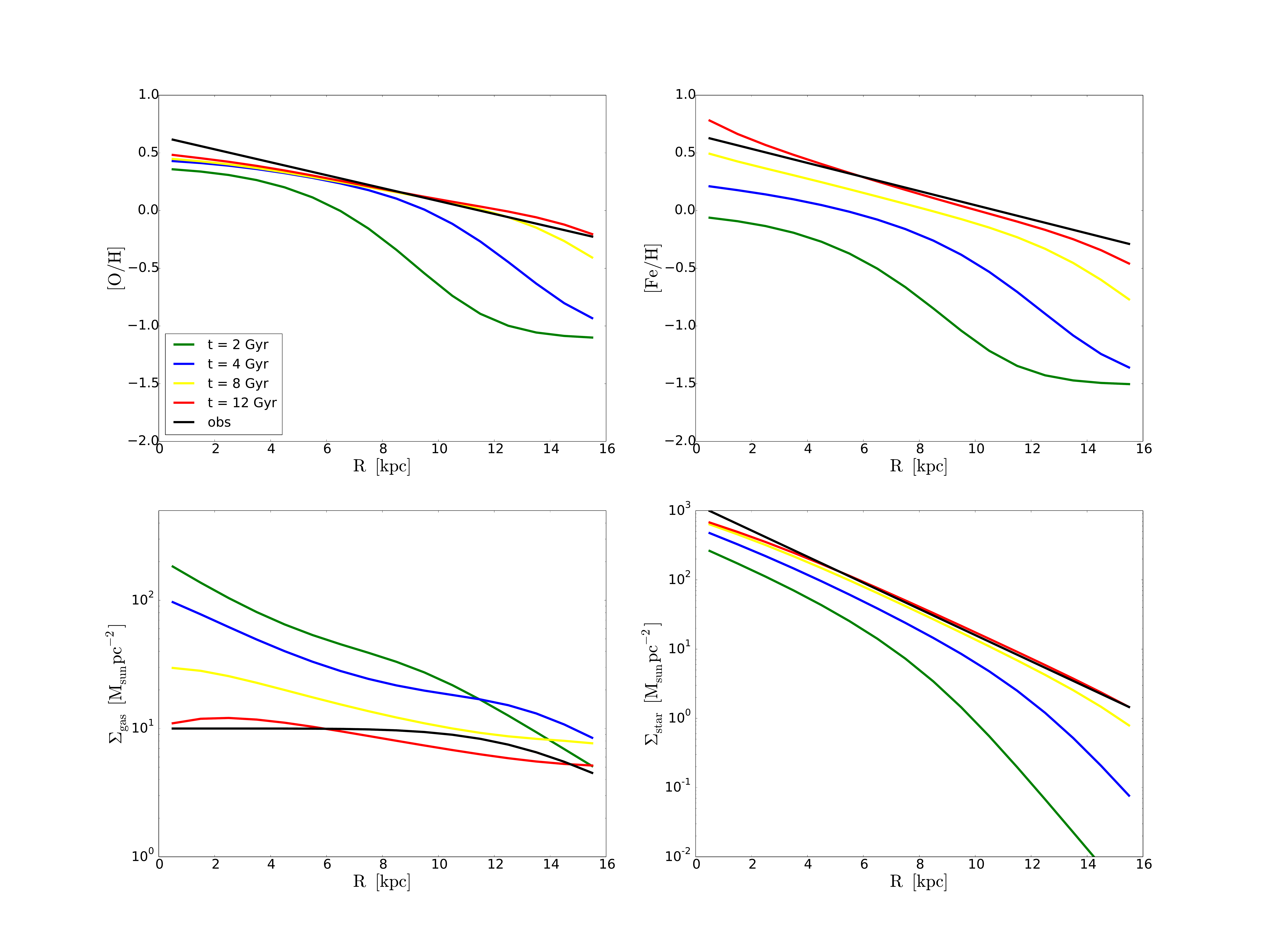}
\end{center}
\caption{The radial profiles of surface densities for [O/H], [Fe/H], gas, and stars obtained in the NRM model with the best fit parameter set. The green, blue, yellow, and red lines show the results at $t$ = 2, 4, 8, and 12 Gyr, respectively. The black lines are the observed radial profiles.}
\label{fig:rp_nrm} 
\end{figure}

\begin{figure}
\begin{center}
\includegraphics[width=17cm,height=12cm]{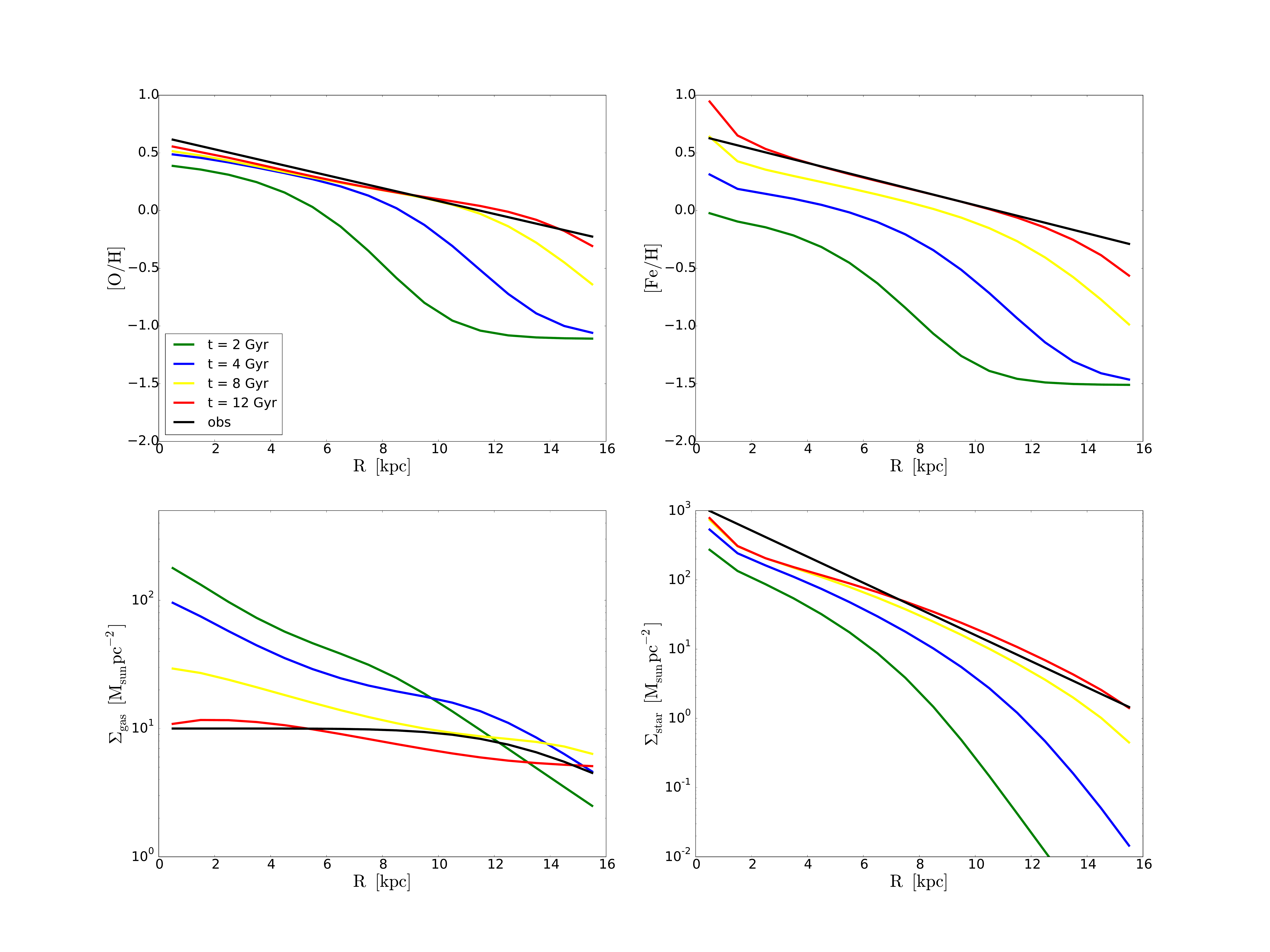}
\end{center}
\caption{The same as Figure \ref{fig:rp_nrm}, but for the CRM model.}
\label{fig:rp_crm} 
\end{figure}

\begin{figure}
\begin{center}
\includegraphics[width=17cm,height=12cm]{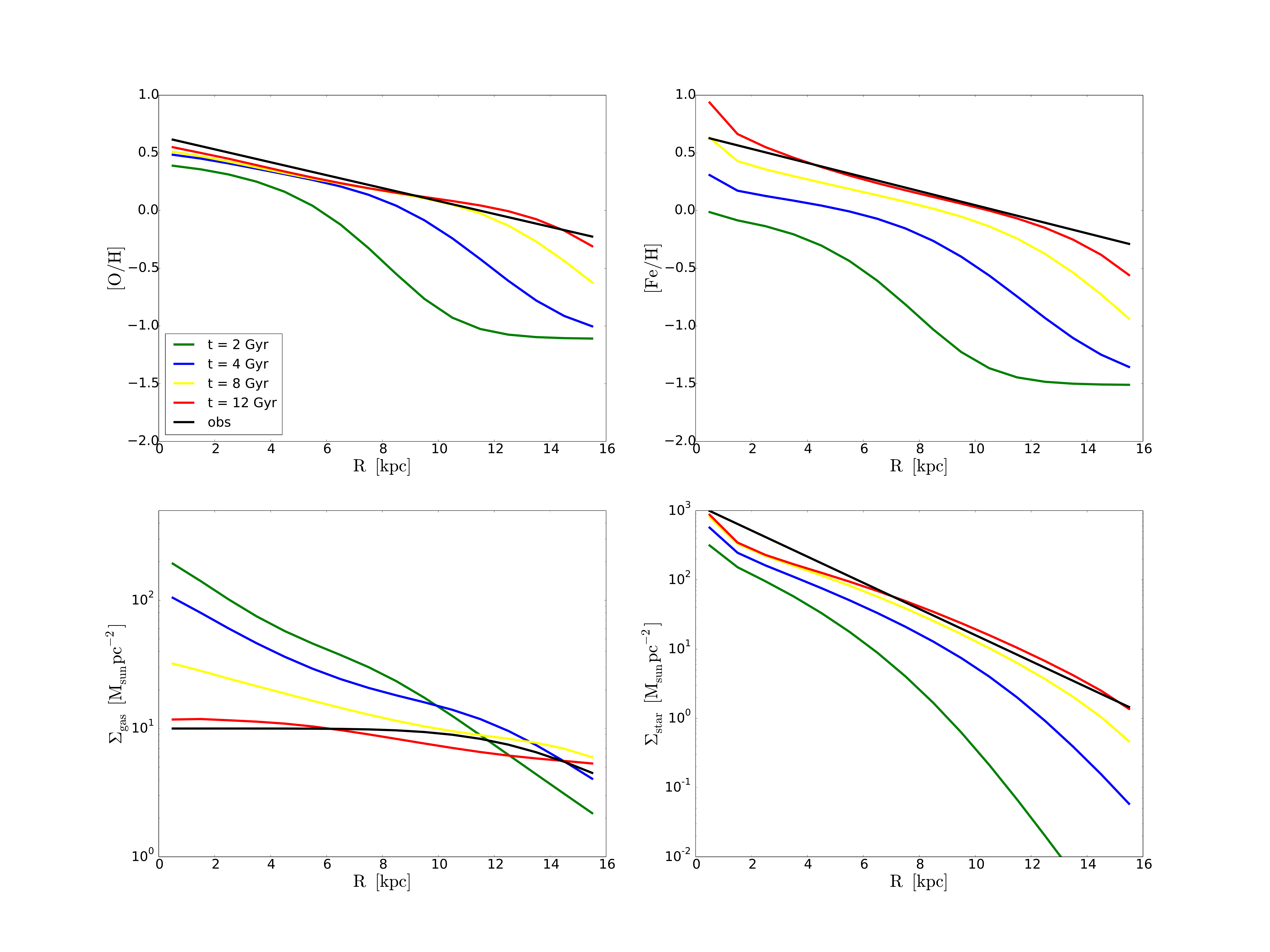}
\end{center}
\caption{The same as Figure \ref{fig:rp_nrm}, but for the DRM model.}
\label{fig:rp_drm} 
\end{figure}

Figure \ref{fig:rp_nrm}, \ref{fig:rp_crm} and \ref{fig:rp_drm} show the time evolution of the radial profiles of gas, star, [O/H] and [Fe/H] obtained from the calculation based on the set of best-fit parameters for the NRM, CRM and DRM models, respectively. The green, blue, yellow and red lines in each panel in these figures represent the results at $t$ =  2, 4, 8, and 12 Gyr, respectively, and the black line denotes the observed profiles. From comparison between the red and black lines in these figures, we find that all of our three models can give the moderately good reproduction for the present chemo-structural properties of the Galactic disk. Thus our chemical evolution models are appropriate in comparison of the Galactic stellar disk. In the next section, we present the detailed properties of chemical evolutions provided by these best models with three different assumptions for radial migration.

%%% Sec.3 %%%%%%%%%%%%%%%%%%%%%%%%%%%%%%%%%%%%%%%%%%%%%%%%
\section{RESULTS}
In this section, by investigating the results of the three radial migration models, we study the difference in the stellar distribution on the [$\alpha$/Fe]-[Fe/H] diagram, arising from the different radial migration histories. 

\begin{figure}
\begin{center}

\includegraphics[width=18cm,height=5cm]{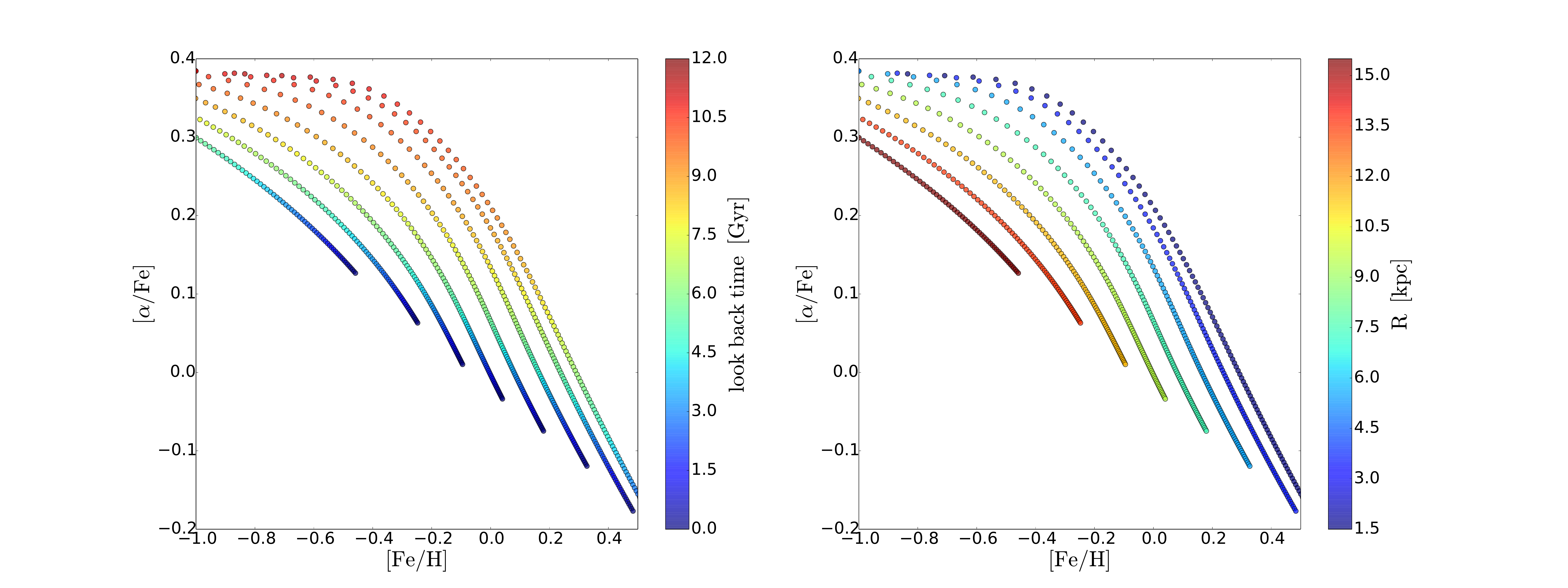}
\\
\includegraphics[width=18cm,height=5cm]{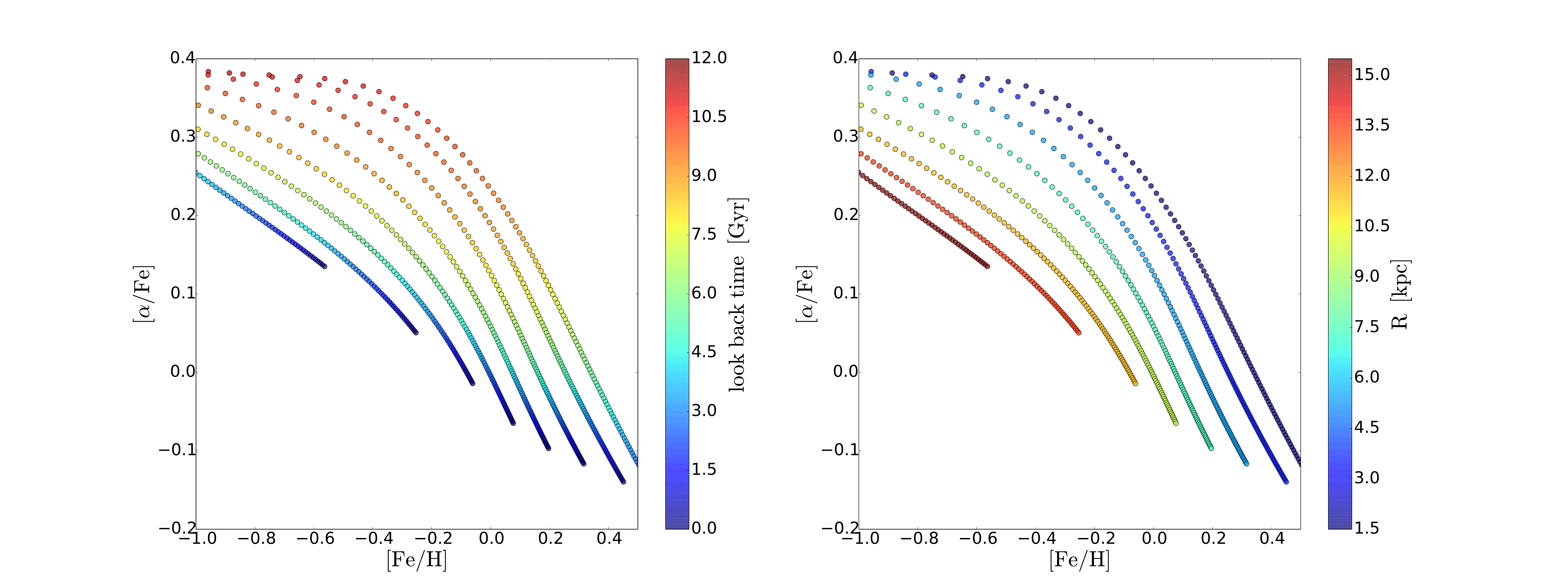}
\\
\includegraphics[width=18cm,height=5cm]{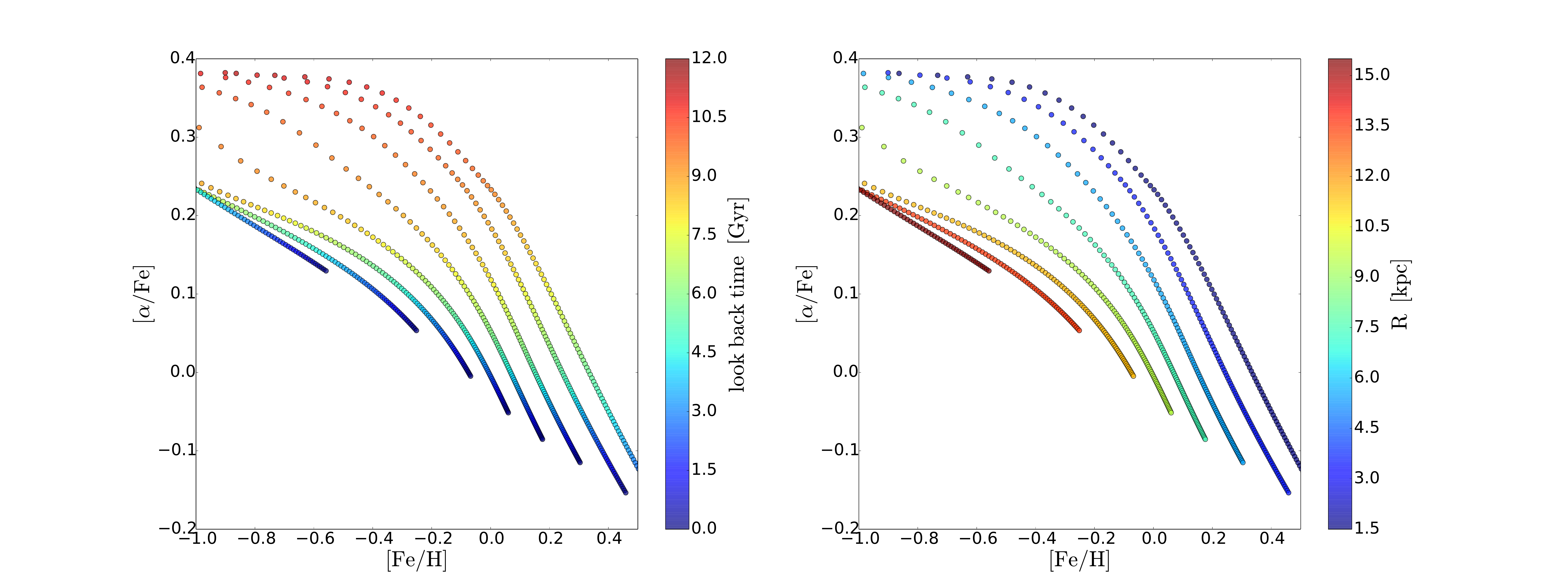}

\end{center}
\caption{Each plot in the top, middle, and bottom panels show the [O/Fe] and [Fe/H] of gas in each radial ring at each time step for the NRM, CRM, and DRM models, respectively. The color of each plot in the left and right panels for each model corresponds to the look back time and the radius, respectively.}
\label{fig:plot}
\end{figure}

In the top, middle, and bottom panels of Figure \ref{fig:plot}, we plot the [$\alpha$/Fe] vs. [Fe/H] for gas in each radial ring at each time step for the NRM, CRM, and DRM models, respectively. The color of each plot in the left and right panels shows the look back time and the radius, respectively. These panels show a general trend that at a fixed [Fe/H] disk gas at an earlier epoch has a higher [$\alpha$/Fe] ratio (left panel) and that at a fixed [$\alpha$/Fe] inner disk gas has a higher [Fe/H] (right panel) for all the radial migration models. There are also some remarkable differences between the three models. Comparing the CRM with NRM models, the former model shows a more rapid decrease of [$\alpha$/Fe] with time at a large $R$ than the latter model. This difference seen at a larger $R$ is due to the increase of the number ratio of SNe Ia to SNe II, $N_\mr{Ia}/N_\mr{II}$, caused by radial migration in which intermediate and old disk stars, which eventually explode as SNe Ia, migrate from inner to outer disk regions. However, since this migration event of stars proceeds gradually in this CRM model, the increase of $N_\mr{Ia}/N_\mr{II}$ remains insignificant that the time dependence of [$\alpha$/Fe]-[Fe/H] diagram at each radius does not change so dramatically. 

On the other hand, the DRM model shows a remarkable property compared with other models. In this model, [$\alpha$/Fe] at $R \gtrsim 6$ kpc decreases rapidly at the look back time of $\sim$ 10 Gyr, corresponding to the timing of the discontinuous radial migration event, so that the density of plotted dots on the [$\alpha$/Fe]-[Fe/H] plane is made sparse near [$\alpha$/Fe] $\sim$ 0.2-0.3 and [Fe/H] $\sim$ $-$0.5. In contrast, in the inner disk regions of $R \lesssim 3$ kpc the declines of [$\alpha$/Fe] temporarily suspend immediately after the discontinuous radial migration event, and consequently the plotted dots on the plane are crowded around [$\alpha$/Fe] $\sim$ 0.2-0.3 and [Fe/H] $\sim$ 0. These changes in the inner and outer disk regions are explained by the rapid decrease and increase of $N_\mr{Ia}/N_\mr{II}$, respectively, arising from the rapid transfer of disk stars from inner to outer disk regions, which occurs much rapidly in the DRM model. Thus the event of fast radial redistribution of disk stars as supposed in the DRM model provides a remarkable feature in the [$\alpha$/Fe]-[Fe/H] diagram for disk gas.

\begin{figure}
\begin{center}

\includegraphics[width=18cm,height=4cm]{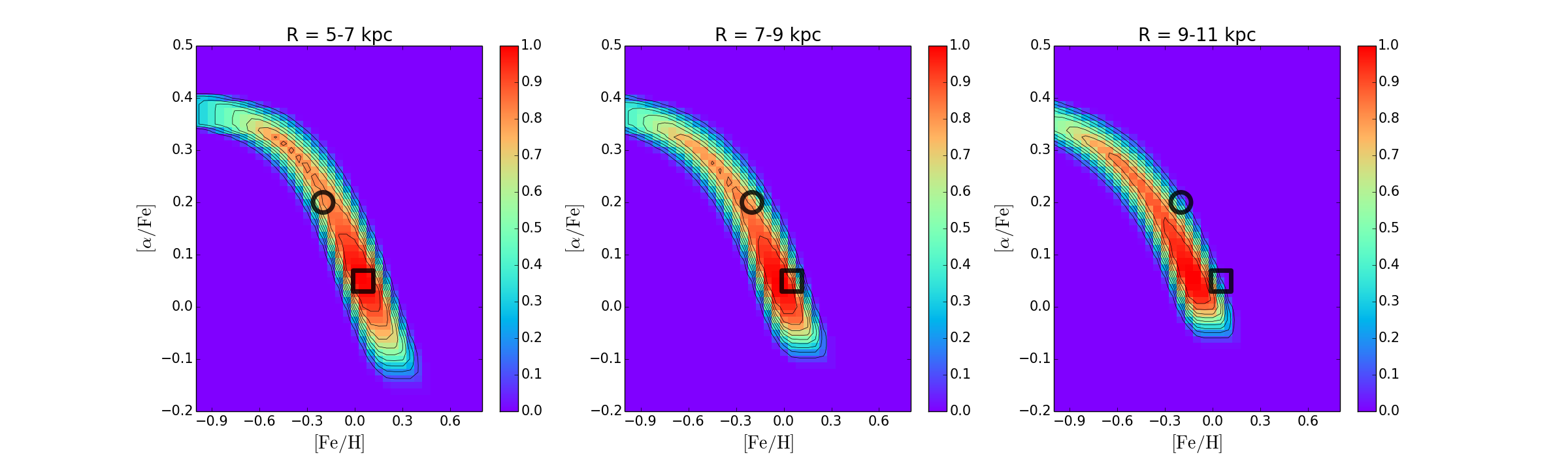}
\\
\includegraphics[width=18cm,height=4cm]{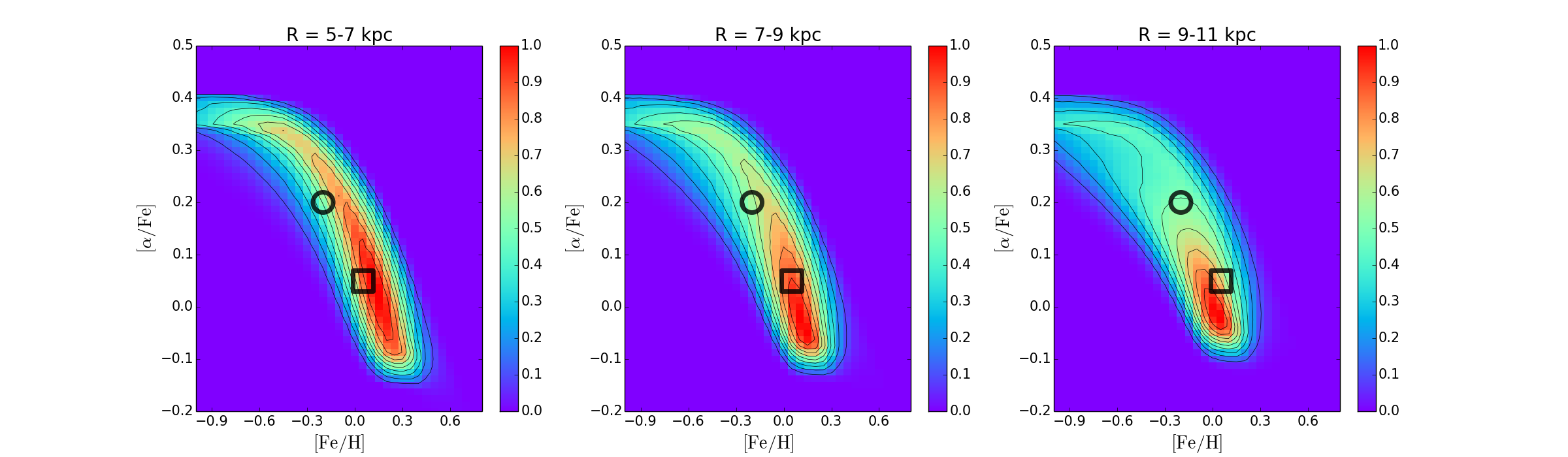}
\\
\includegraphics[width=18cm,height=4cm]{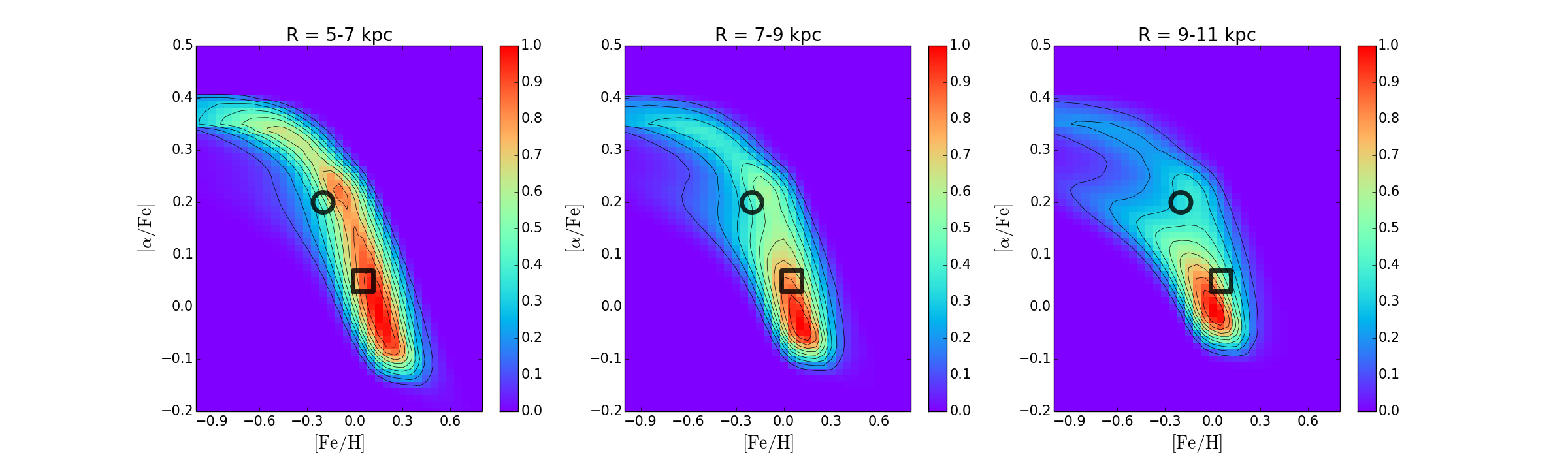}

\end{center}
\caption{The contours and color maps in the top, middle, and bottom panels represent the stellar distributions on the [O/Fe]-[Fe/H] plane at $t$ = 12 Gyr, reproduced in the NRM, CRM, and DRM models, respectively. The left, middle, and right panels for each model show the stellar distribution observed at the inner ($R$ = 5-7 kpc), solar neighborhood ($R$ = 7-9 kpc), and the outer disk region ($R$ = 9-11 kpc), respectively. The stellar density in each panel is normalized that the maximum density on the panel is unity. The black circle and square at ([$\alpha$/Fe], [Fe/H]) = (0.2, $-$0.2) and (0.05, 0.05) in each panel indicate the [$\alpha$/Fe] and [Fe/H] values of the observed high and low-$\alpha$ peaks for the nearby disk stars, respectively.}
\label{fig:dist}
\end{figure}

To specify how these differences in the chemical evolution of disk gas shown in Figure \ref{fig:plot} appear in the observed disk stars, we show the present stellar distributions on the [$\alpha$/Fe]-[Fe/H] plane with color maps and contours  in Figure \ref{fig:dist}. The results of the NRM, CRM, and DRM models are shown in the top, middle, and bottom panels, respectively. The left, middle, and right panels for each model represent the stellar distribution observed at the inner ($R$ = 5-7 kpc), solar neighborhood ($R$ = 7-9 kpc), and the outer disk region ($R$ = 9-11 kpc), respectively. For comparison with observations, we plot the black circle and square at ([$\alpha$/Fe], [Fe/H]) = (0.2, $-$0.2) and (0.05, 0.05) in each panel, which delineate the observed [$\alpha$/Fe] and [Fe/H] values of the high and low-$\alpha$ peaks for the nearby disk stars, respectively (see Figure 3 in Hayden et al. 2015).

As shown in many previous works, the effect of radial migration of stars makes their distribution on the [$\alpha$/Fe]-[Fe/H] plane diffuse at all radial ranges. In the CRM model, we find that the distribution of stars in this abundance-ratio diagram is unimodal and that the metallicity of their density peak remains higher at inner $R$, suggesting the negative radial metallicity gradient as also seen in the NRM model. In contrast, it is remarkable that at all radii the DRM model produces bimodal stellar density distributions on the [$\alpha$/Fe]-[Fe/H] plane, where the [$\alpha$/Fe] ratios of the high and low-$\alpha$ density peaks are $\sim$ 0.2-0.3 and $\sim$ $-$0.05, respectively. The stars located at high-[$\alpha$/Fe] density peaks are originally born in inner disk regions and have moved outward. In addition, as a result of the discontinuous migration event, the number density of stars near  [$\alpha$/Fe] $\sim$ 0.2-0.3 and [Fe/H] $\sim$ $-$0.5, is made sparse related to the same effect seen in the bottom panels of Figure \ref{fig:dist}, so that the bimodal distributions as observed in the Milky Way disk stars are successfully reproduced in the DRM model. It is also worth remarking that while the metallicity of the high-[$\alpha$/Fe] peak is generally independent of radius, that of the low-[$\alpha$/Fe] peak decreases with increasing radius, in good agreement with the recent observational results for the Galactic stellar disk (Hayden et al. 2015; Kordopatis et al. 2015). 

The comparisons with the observed high- and low-[$\alpha$/Fe] peaks denoted with the black circle and square in the figure show that in our model the metallicity of the high-[$\alpha$/Fe] peak and the [$\alpha$/Fe] ratio of the low-[$\alpha$/Fe] peak are somewhat higher and lower than the observed values, respectively. This small difference from the observational data is due to our simple constraints adopted in the MCMC procedure, namely based on only the present radial profiles of gas, stars and metallicity of the Milky Way disk, without using the distributions of [$\alpha$/Fe] and [Fe/H] as constraints. More refined modelings and extensive searches of model parameters are needed to reproduce the details of the observed abundance distributions, which is however beyond the scope of this work, aiming to demonstrate the impact of the radial migration history on the [$\alpha$/Fe] vs. [Fe/H] diagram.

Thus our model experiments suggest that the discontinuous radial migration history can be a candidate solution in explaining the observed bimodal distribution of the Milky Way disk stars on the [$\alpha$/Fe]-[Fe/H] plane. In the next section, we discuss the possibility of such a discontinuous radial migration of stars as represented in the DRM model in more detail and examine its significance in the chemical evolution history of the Galactic stellar disk.

%%% Sec.4 %%%%%%%%%%%%%%%%%%%%%%%%%%%%%%%%%%%%%%%%%%%%%%%%
\section{DISCUSSION}
%%% Sec.4.1 %%%%%%%%%%%%%%%%%%%%%%%%%%%%%%%%%%%%%%%%%%%%%%%%
\subsection{Dependence of the model results on $\trm$ and $\srmz$}
A discontinuous radial migration event is characterized by the value of $\trm$ and $\srmz$, corresponding to the timing and the magnitude of the radial migration. We here examine how different choices of these value can affect the stellar distribution on the [$\alpha$/Fe]-[Fe/H] plane. 

\begin{figure}
\begin{center}

\includegraphics[width=18cm,height=4cm]{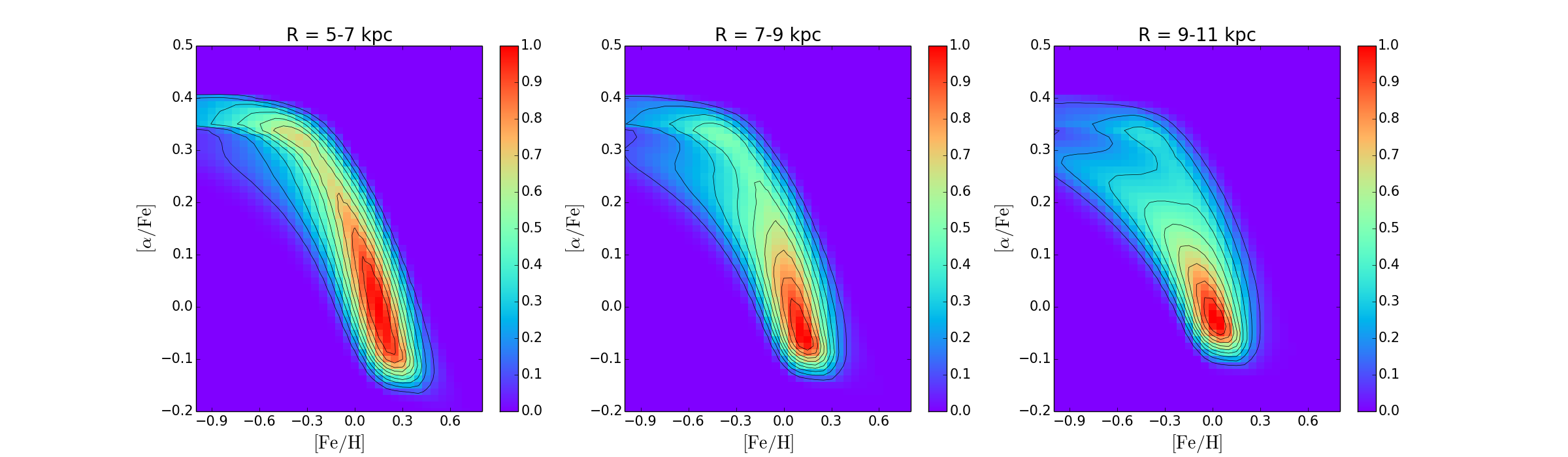}
\\
\includegraphics[width=18cm,height=4cm]{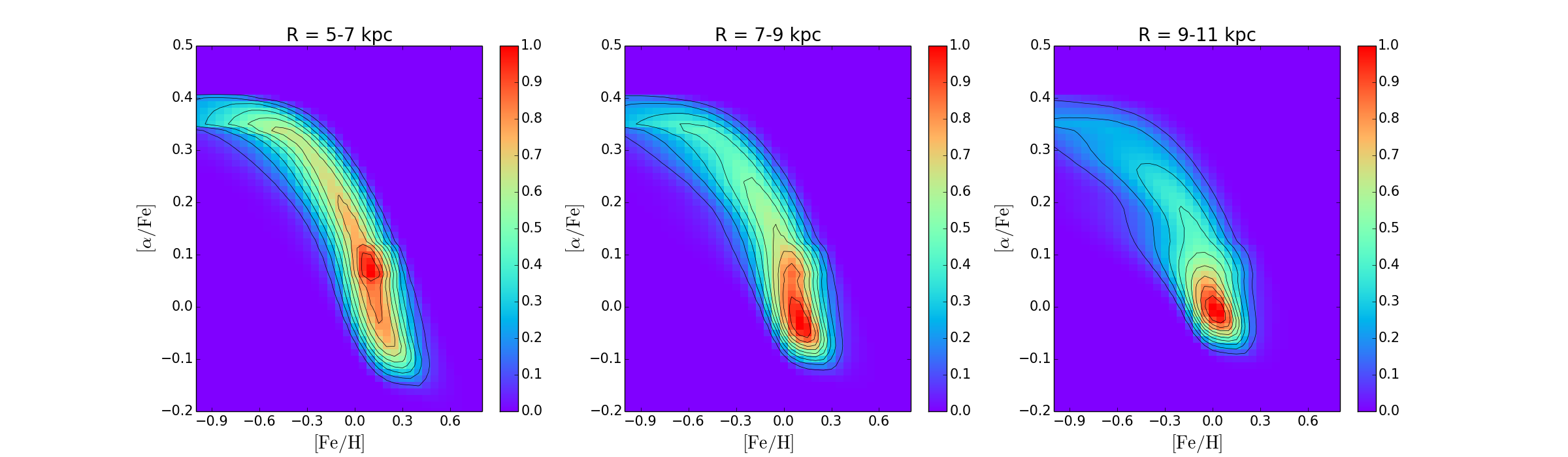}

\end{center}
\caption{The upper and lower panels represent the stellar density distribution on the [O/Fe]-[Fe/H] plane for the two DRM models with $\trm$ = 1, and 4 Gyr, respectively, instead of $\trm$ = 2 Gyr assumed in the fiducial model in the bottom panels of Figure \ref{fig:dist}. For both models we set $\srmz$ = 3 kpc in agreement with the fiducial model. The left, middle, and right panels for each model show the stellar distribution observed at the inner ($R$ = 5-7 kpc), solar neighborhood ($R$ = 7-9 kpc), and the outer disk region ($R$ = 9-11 kpc), respectively. The stellar density in each panel is normalized that the maximum density on the panel is unity.}
\label{fig:trm}
\end{figure}

Figure \ref{fig:trm} shows the stellar density distribution on the [$\alpha$/Fe]-[Fe/H] plane for the two DRM models with different values of $\trm$ from the fiducial DRM model of $\trm$ = 2 Gyr. The top and bottom panels are the results of the model with $\trm$ = 1 and 4 Gyr, respectively, for which we set $\srmz$ = 3 kpc for both models. 

We find from the top panels that the model with a smaller value of $\trm$ does not produce a clear bimodal distribution. This may be explained by the too early timing of the discontinuous radial migration event, in which the amount of disk stars that are transferred to the outer disk region is not enough to cause a rapid increase of $N_\mr{Ia}/N_\mr{II}$: the mass density of the stellar disk was still small at such an early epoch. On the other hand, although the model with $\trm$ = 4 Gyr seems to reproduce a bimodal distribution, the two peaks are mostly merged because the [$\alpha$/Fe] ratio in the inner disk region is made already low before the radial migration event occurs. Thus the formation of such a bimodal distribution is sensitive to the timing of the discontinuous radial migration event: to reproduce the observed bimodality in our Galactic stellar disk, $\trm$ is confined to be around 2-3 Gyr.

\begin{figure}
\begin{center}

\includegraphics[width=18cm,height=4cm]{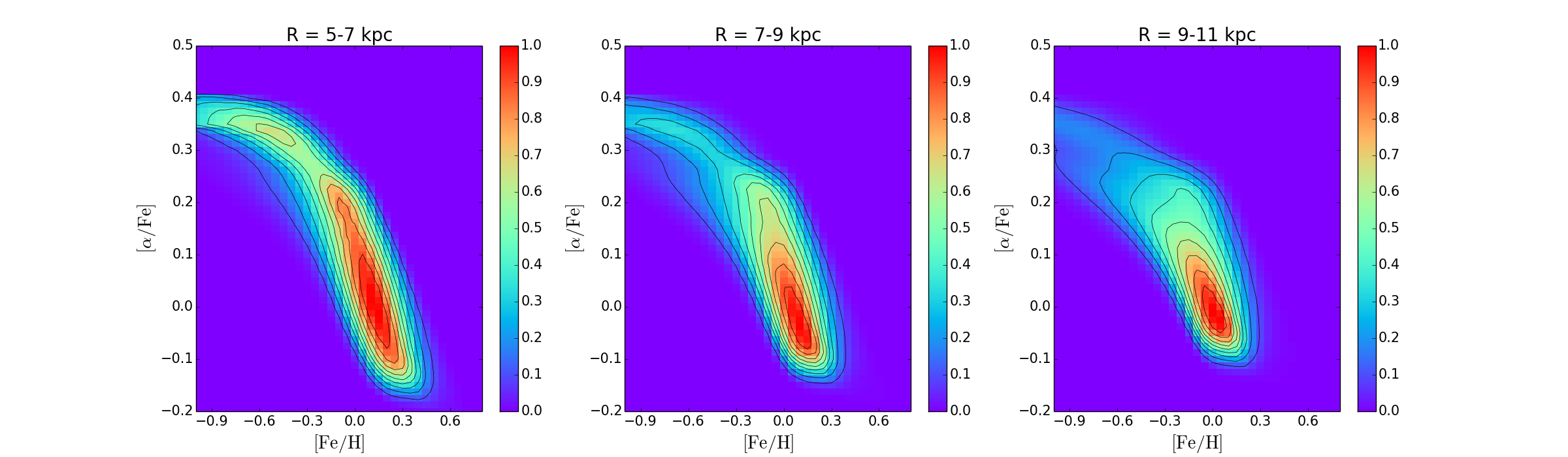}
\\
\includegraphics[width=18cm,height=4cm]{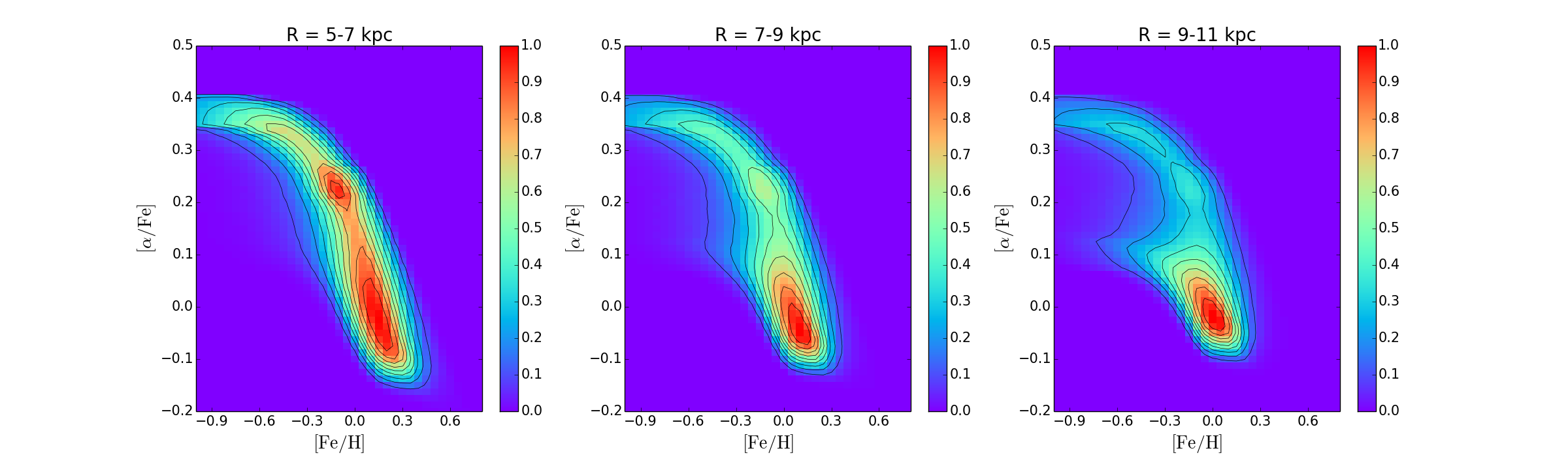}

\end{center}
\caption{The upper and lower panels represent the stellar density distribution on the [O/Fe]-[Fe/H] plane for the two DRM models with $\srmz$ = 2, and 4 kpc, respectively, instead of $\srmz$ = 3 kpc assumed in the fiducial model in the bottom panels of Figure \ref{fig:dist}. For both models we set $\trm$ = 2 Gyr in agreement with the fiducial model. The left, middle, and right panels for each model show the stellar distribution observed at the inner ($R$ = 5-7 kpc), solar neighborhood ($R$ = 7-9 kpc), and the outer disk region ($R$ = 9-11 kpc), respectively. The stellar density in each panel is normalized that the maximum density on the panel is unity.}
\label{fig:srmz}
\end{figure}

Next, we investigate the dependence of the model results on $\srmz$. Figure \ref{fig:srmz} shows the results for the DRM model when we adopt two different values of $\srmz$. The top and bottom panels correspond to the cases of $\srmz$ = 2 and 4 kpc, respectively, where we set $\trm$ = 2 Gyr for both models. It follows that while the high-$\alpha$ peak around [$\alpha$/Fe] $\sim$ 0.2-0.3 disappears in the model with $\srmz$ = 2 kpc, it remains in the model with $\srmz$ = 4 kpc but more clearly than in the fiducial DRM model with $\srmz$ = 3 kpc. Such a dependence on $\srmz$ is simply explained in terms of the increase of the fraction of the inner disk stars with increasing $\srmz$, which can reach the outer disk region by radial migration. Thus, the value of $\srmz$, namely the magnitude of the radial migration event, is also an important ingredient for the formation of the bimodal distribution as well as $\trm$: to produce a clear bimodal distribution we need $\srmz \geq$ 3 kpc, where its actual value can be derived from the number ratio between stars belonging to the high and low-$\alpha$ sequences.

Based on the above experiments, as one of possible mechanisms to produce the observed bimodal distribution of the disk stars on the [$\alpha$/Fe] plane, we propose the past event of a rapid and discontinuous radial migration with a diffusion scale of $\geq$ 3 kpc which may have occurred during 2-3 Gyr after the onset of the disk formation. In the next section we discuss the possibility of this scenario in the context of  galaxy evolution.

%%% Sec.4.2 %%%%%%%%%%%%%%%%%%%%%%%%%%%%%%%%%%%%%%%%%%%%%%%%
\subsection{What would cause discontinuous radial migration?}
Here, we consider the possibility that the discontinuous radial migration event is triggered by a minor merger of a relatively massive satellite galaxy onto the Milky Way stellar disk. Following the above experiments, this last minor merger event, which can affect the dynamical structure of the disk, may have occurred at the look back time of 9-10 Gyr. This scenario may not be unrealistic in the cosmological context: according to Ruiz-Lara et al. (2016), in disk galaxies like the Milky Way, last minor mergers with satellite galaxies being more massive than 1/10 virial mass of host galaxies can take place at a look back time of $\sim$ 9 Gyr on average. 

\begin{figure}
\begin{center}
\includegraphics[width=17cm,height=12cm]{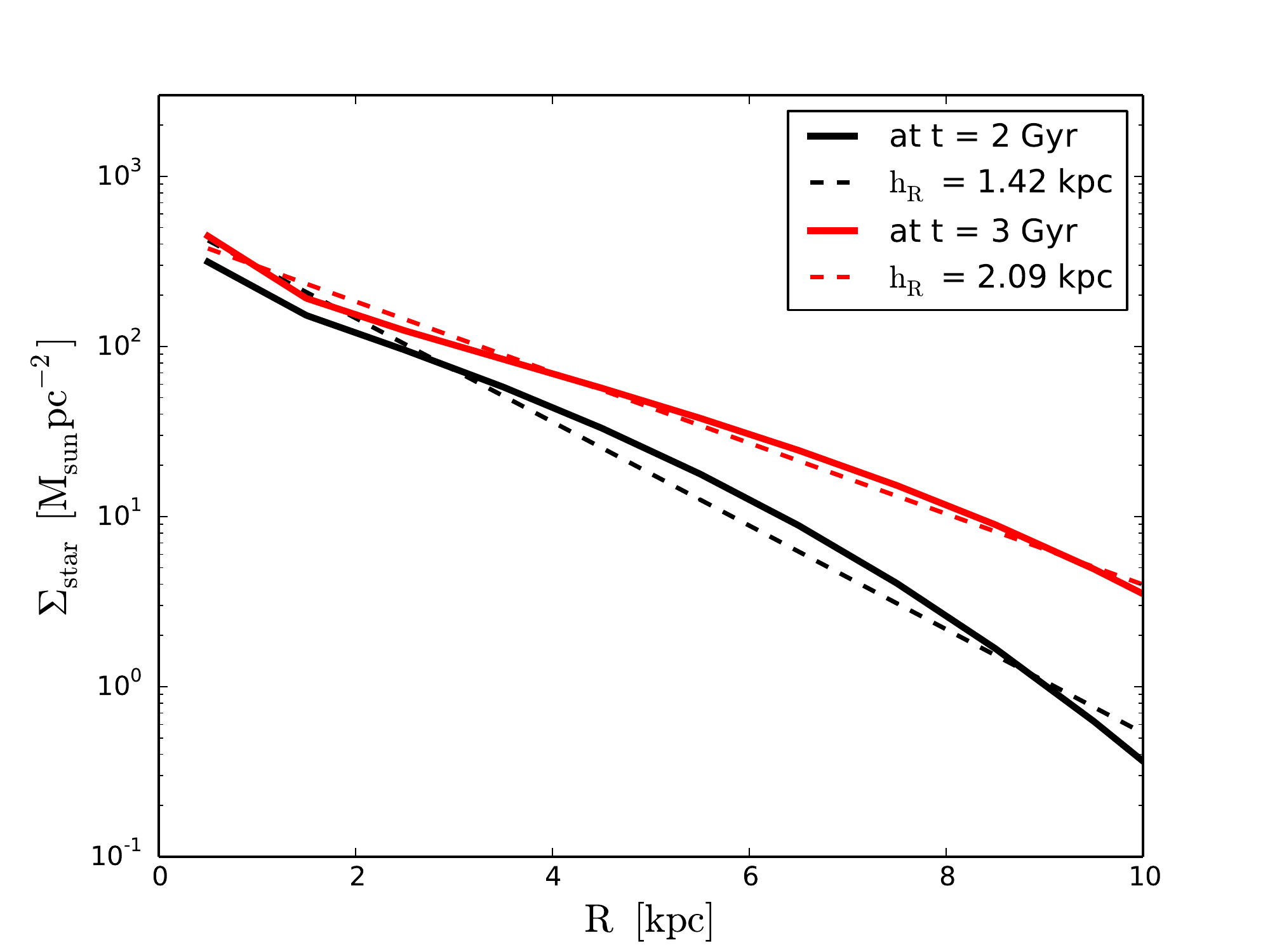}
\end{center}
\caption{The black and red solid lines show the stellar density profiles for the fiducial DRM model at $t$ = 2 and 3 Gyr, corresponding to immediately before and after the discontinuous radial migration event, respectively. The dashed lines represent the results of exponential fitting to these stellar density profiles in the radial range of $R \leq$ 10 kpc, and the scale lengths obtained from these fittings are also shown in the upper-right rectangle of the panel.}
\label{fig:hr} 
\end{figure}

On the other hand, another condition of $\srmz \geq$ 3 kpc seems rather specific when we consider the possibility of vastly various properties of minor mergers. Therefore it is worth assessing the possibility of such a discontinuous radial migration event in terms of the evolution of a scale length of stellar disk rather than $\srmz$. In Figure \ref{fig:hr} the black and red solid lines show the stellar density profiles for the fiducial DRM model at $t$ = 2 and 3 Gyr, corresponding to immediately before and after the discontinuous radial migration event, respectively. The dashed lines represent the results of the exponential fitting to these stellar density profiles in the radial range of $R \leq$ 10 kpc, and the scale lengths obtained from these fittings are also shown in the upper-right rectangle in Figure \ref{fig:hr}. This figure implies that the discontinuous radial migration event with $\srmz$ = 3 kpc increases the scale length of the stellar disk from 1.42 kpc to 2.09 kpc. Such a change of the scale length is reasonably achievable in the minor merger of a satellite and associated disk heating. Indeed, the N-body simulations in Villalobos \& Helmi (2008) show that the scale length of a disk galaxy, in which the stellar mass is similar to that of our model galaxy at the time of the rapid radial migration event, is grown by minor merger heating from 1.65 kpc to 2.1 kpc. Thus the discontinuous radial migration assumed in our DRM model may be the case in a minor merger event between massive a satellites and the Milky Way stellar disk.

We note here that such a minor merger event may also trigger not only discontinuous radial migration but also discontinuous disk thickening. Bovy et al. (2012) investigated the spatial distribution of the Galactic mono-abundance disk populations and suggested that the vertical structure of the Galactic stellar disk has evolved continuously rather than discontinuously. This implies that a supposed merger event does not increase the disk thickness so dramatically, although this may be unlikely in light of the associated dynamical heating of a stellar disk. To resolve this paradoxical situation, one possible solution is to rely on a special minor merger event, in which the orbit of an accreting satellite is almost parallel to the stellar disk plane of the host galaxy. Villalobos \& Helmi (2008) show that a minor merger event of a massive satellite with a parallel orbit to the initial stellar disk can thicken the stellar disk with an amount only to the scale height of $\sim$ 0.6 kpc, which is much thinner than the thickest population of the present Galactic stellar disk. Thus, if such a minor merger event without causing significant disk thickening occurred in the Galactic past, then the issue mentioned above may be resolved.

%%% Sec.5 %%%%%%%%%%%%%%%%%%%%%%%%%%%%%%%%%%%%%%%%%%%%%%%%
\section{Summary \& Conclusion}
In this paper, we have calculated the chemo-dynamical model to investigate the influence of radial migration histories on the chemical evolution of a disk galaxy, in particular focusing on stellar distribution on the [$\alpha$/Fe]-[Fe/H] plane. For this purpose, we have examined the three models, where we consider the models with both continuous and discontinuous radial migration events, in comparison with the no radial migration model. 

We found that radial migration can speed up the evolution of [$\alpha$/Fe] in outer disk regions by increasing the rate of SNe Ia to SNe II because of the net transfer of intermediate and old disk stars, including progenitors of SNe Ia, from inner to outer disk regions. Moreover, in the model of the rapid and discontinuous radial migration, such an effect is stronger than the case of the continuous radial migration, thereby the properties of the [$\alpha$/Fe]-[Fe/H] diagram in the inner and outer disk regions are made distinctly different. Thus chemical evolution histories of disk galaxies can be significantly affected by their radial migration histories.

We also found that the discontinuous radial migration can reproduce the bimodal distribution on the [$\alpha$/Fe]-[Fe/H] plane as observed in the Galactic stellar disk. Our model calculation predicts that the high-[$\alpha$/Fe] sequence consists of disk stars which were originally born in inner disk regions and have been moved to the solar neighborhood by the rapid and discontinuous radial migration event. On the other hand, the low-[$\alpha$/Fe] sequence stars are originated from the outer disk region, where the rapid decrease of [$\alpha$/Fe] caused by the rapid increase of the rate of SN Ia to SN II is achieved. A remarkable point of the bimodal distribution reproduced by our discontinuous radial migration model is that while the metallicity of the high-[$\alpha$/Fe] sequence is roughly independent of radius, that of the low-[$\alpha$/Fe] sequence increases with increasing radius, reflecting the negative radial metallicity gradient, in good agreement with the observational property of the Galactic stellar disk. Therefore, we suggest in this paper that the discontinuous radial migration scenario is a key in understanding the observed bimodal distribution of the Milky Way disk stars on the [$\alpha$/Fe]-[Fe/H] plane.

To further understand this formation scenario of the bimodal distribution, we have investigated the dependence of the bimodality on the model parameters for radial migration. It is found that the reproduction of a bimodal distribution on the [$\alpha$/Fe]-[Fe/H] plane is very sensitive to the timing and magnitude of the discontinuous radial migration event. According to our model calculations, in order to reproduce such a bimodal distribution on the [$\alpha$/Fe] plane observed in the Galactic stellar disk, a rapid and discontinuous radial migration event with the diffusion length of stars of $\geq$ 3 kpc is required to occur during 2-3 Gyr after the onset of the disk formation. Thus the bimodality in the [$\alpha$/Fe]-[Fe/H] distribution of disk stars is a key in elucidating the detailed radial migration history. Further studies based on detailed numerical simulations for minor merging events in the Galactic past may be necessary to obtain their effects on the mechanism of radial migration of stars and the subsequent role in the chemo-dynamical evolution of a disk galaxy like the Milky Way.

%%%%%%%%%%%%%%%%%%%%%%%%%%%%%%%%%%%%%%%%%%%%%%%%%%%%%%%%%
\acknowledgments
We are grateful to Kenji Bekki for his invaluable comments on this work. This work is supported in part by JSPS Grant-in-Aid for Scientific Research (No. 27-2450 for DT) and MEXT Grant-in-Aid for Scientific Research on Innovative Areas (No. 15H05889 for MC).

%%%%%%%%%%%%%%%%%%%%%%%%%%%%%%%%%%%%%%%%%%%%%%%%%%%%%%%%%

\clearpage
%%%%%%%%%%%%%%%%%%%%%%%%%%%%%%%%%%%%%%%%%%%%%%%%%%%%%%%%%

%%%% table %%%%%

%%% Table 1 %%%
\begin{table}
	\begin{center}
		\caption{The list of free parameters in our chemo-dynamical model}
		\begin{tabular}{c|l} \hline\hline
			$\mtotin$ & Total mass of inflowing gas \\
			$\hrin$ & Scale length of inflowing gas \\
			$\tauinz$ & Time scale of gas inflow at $R$ = 0 kpc \\
			$\tauine$ & Time scale of gas inflow at $R$ = 8 kpc \\
			$\alpha$ & Power-law index characterizing the radial dependence of the time scale of gas inflow \\
			$\Lambda_0$ & Mass loading factor at $R$ = 0 kpc \\
			$\Lambda_8$ & Mass loading factor at $R$ = 8 kpc \\
			$\beta$ & Power-law index characterizing the radial dependence of mass loading factor \\
			$\fia$ & Parameter controlling the rate of Type Ia supernova \\			
			 \hline
		\end{tabular}
		\label{tb:intr} 
	\end{center}
\end{table}

%%% Table 2 %%%
\begin{table}
	\begin{center}
		\caption{The results of the estimation of free parameters for the three radial migration models}
		\begin{tabular}{c|rr|rr|rr} \hline\hline
			\multicolumn{1}{c|}{ } & \multicolumn{2}{c|}{NRM} & \multicolumn{2}{c|}{CRM} & \multicolumn{2}{c}{DRM} \\ 
			\cline{2-7}
			 & best\tablenotemark{a} & median\tablenotemark{b} & best\tablenotemark{a} & median\tablenotemark{b} & best\tablenotemark{a} & median\tablenotemark{b} \\ \hline
			$\mtotin$ [$10^{10} \ M_\odot$] & 12.5 & 12.36$^{+1.67}_{-1.39}$ & 9.23 & 9.57$^{+1.29}_{-1.07}$ & 9.7 & 10.5$^{+1.52}_{-1.32}$ \\
			$\hrin$ [kpc] & 3.44 & 3.51$^{+0.29}_{-0.25}$ & 3.18 & 3.15$^{+0.34}_{-0.27}$ & 3.09 & 2.98$^{+0.29}_{-0.23}$ \\
			$\tauinz$ [Gyr] & 2.66 & 2.9$^{+0.55}_{-0.99}$ & 2.85 & 3.21$^{+0.38}_{-0.39}$ & 2.91 & 3.13$^{+0.35}_{-0.37}$ \\
			$\tauine$ [Gyr] & 5.75 & 6.12$^{+1.34}_{-0.83}$ & 7.18 & 9.91$^{+3.94}_{-2.09}$ & 8.28 & 9.33$^{+5.02}_{-2.15}$ \\
			$\alpha$ & 0.98 & 1.2$^{+0.77}_{-0.58}$ & 1.64 & 2.76$^{+1.46}_{-1.0}$ & 1.98 & 2.84$^{+1.73}_{-1.03}$ \\
			$\Lambda_0$ & 0.63 & 0.73$^{+0.23}_{-0.18}$ & 0.49 & 0.55$^{+0.24}_{-0.29}$ & 0.49 & 0.57$^{+0.28}_{-0.4}$ \\
			$\Lambda_8$ & 1.54 & 1.58$^{+0.22}_{-0.19}$ & 1.58 & 1.63$^{+0.26}_{-0.22}$ & 1.53 & 1.64$^{+0.25}_{-0.21}$ \\
			$\beta$ & 2.18 & 2.38$^{+0.58}_{-0.51}$ & 1.73 & 1.65$^{+0.72}_{-0.71}$ & 1.55 & 1.51$^{+0.78}_{-0.76}$ \\
			$\fia$ [$10^{-3}$] & 0.8 & 0.88$^{+0.19}_{-0.16}$ & 0.85 & 1.01$^{+0.22}_{-0.17}$ & 0.87 & 0.98$^{+0.21}_{-0.18}$ \\
			\hline
		\end{tabular}
		\tablenotetext{a}{The best value obtained from the MCMC estimation.}
		\tablenotetext{b}{The median value obtained from the MCMC estimation and the sub- and super-scripts represent the difference from the 16 and 84 percentile value, respectively.}
		\label{tb:intr} 
	\end{center}
\end{table}

\clearpage

%%%% figure %%%%%

%%% Figure 1 %%%

%%% Figure 2 %%%

%%% Figure 3 %%%

%%% Figure 4 %%%

%%% Figure 5 %%%

%%% Figure 6 %%%

%%% Figure 7 %%%

%%% Figure 8 %%%

%%% Figure 9 %%%

%%% Figure 10 %%%

%%% Figure 11 %%%

%%% Figure 12 %%%

\end{document}